\documentclass[traditabstract]{aa}  
\usepackage{natbib}
\def\lunits{$\rm erg\,s^{-1}$~}

\def\xmm{{\it XMM-Newton~}}
\def\chandra{{\it Chandra~}} 
\def\lxl6{{$\rm L_X/L_{6\,\mu m}~$}}

\usepackage{graphicx}
\begin{document}
   \title{Constraining the fraction of Compton-thick AGN in the Universe by modelling the diffuse X-ray background spectrum}


\titlerunning{X-ray background models}
    \authorrunning{A. Akylas et al.}

   \author{A. Akylas
          \inst{1},
          A. Georgakakis
	  \inst{2,1},
	  I. Georgantopoulos
	  \inst{3,1},          
          M. Brightman
          \inst{2}
          \and
          K. Nandra
           \inst{2}
          }

   \institute{Institute of Astronomy \& Astrophysics, National Observatory of Athens, I. Metaxa \& V. Pavlou 1, 15236, Penteli, Greece \\
              \and
               Max-Planck-Institut f\"ur extraterrestrische Physik, Giessenbachstrasse 1, D-85748, Garching bei M\"unchen, Germany\\
              \and 
              INAF-Osservatorio Astronomico di Bologna, Via Ranzani 1, 40127, Italy \\
             }

   \date{Received; accepted}

 \abstract{This paper investigates which  constraints can be placed on
   the fraction  of Compton-thick AGN  in the Universe  from modelling
   the spectrum of  the diffuse X-ray background (XRB).   We present a
   model for the synthesis of the  XRB that uses as input a library of
   AGN X-ray  spectra generated by  Monte Carlo simulations.   This is
   essential to account for the Compton scattering of X-ray photons in
   a dense  medium and the  impact of that  process on the  spectra of
   heavily obscured  AGN.  We identify a  small number  of input
     parameters to
   the  XRB  synthesis code  that  encapsulate  the  minimum level  of
   uncertainty  in reconstructing  the  XRB spectrum.   These are  the
   power-law  index  and high-energy  cutoff  of  the intrinsic  X-ray
   spectra  of AGN,  the  level  of the  reflection  component in  AGN
   spectra, and the fraction of  Compton-thick AGN in the Universe. We
   then map  the volume  of the space  allowed to these  parameters by
   current  observational determinations  of the  XRB spectrum  in the
   range 3-100\,keV.   One of the least-constrained  parameters is the
   fraction  of Compton-thick AGN.   Statistically acceptable  fits to
   the XRB spectrum  at the 68\% confidence level  can be obtained for
   Compton-thick AGN  fractions in the range 5-50\%.   This is because
   of degeneracies  among input parameters  to the XRB  synthesis code
   and uncertainties in  the modelling of AGN spectra  (e.g.  level of
   reflection  fraction).  The most  promising route  for constraining
   the fraction of Compton-thick AGN in the Universe is via the direct
   detection of those sources  in high-energy ($\ga 10$\,keV) surveys.
   We show  that the {\it observed} fraction  of Compton-thick sources
   identified in  the {\it SWIFT/BAT} serendipitous  survey limits the
   {\it  intrinsic} fraction  of Compton-thick  AGN, at  least  at low
   redshift, to  10-20\% (68\% confidence  level).  We  also make
     predictions on  the number density of  Compton-thick sources that
   current  and  future  X-ray  missions  are  expected  to  discover.
   Testing those predictions with data will place tight constraints on
   the  intrinsic  fraction of  Compton-thick  AGN  as  a function  of
   redshift.  }

   \maketitle

\section{Introduction}

The  origin of  the diffuse  X-ray background  (XRB) has  been  at the
centre  of  high-energy  astrophysical  research since  its  discovery
\citep{Giacconi1962}.   Already  by   the  early  90s  numerous  X-ray
experiments have constrained the most salient spectral features of the
XRB,  the power-law  spectral  energy distribution  with photon  index
$\Gamma=1.4$  in  the  energy  interval  2-10\,keV  and  the  peak  at
20-30\,keV  \citep[e.g.][]{Marshall1980,  Gruber1999}.   At  the  same
time, observations  at energies below about  10\,keV demonstrated that
at least a fraction of the XRB intensity is composed of point sources,
some    of    which    were    identified   with    broad-line    QSOs
\citep[Quasi-Stellar  Objects,  e.g.][]{Shanks1991}.   However,  early
suggestions that  the XRB  is the result  of the superposition  of AGN
(Active Galactic Nuclei) across  cosmic time stumbled upon the steeper
X-ray  spectra of  UV-bright  (unobscured) QSOs  compared  to the  XRB
spectrum  below  10\,keV   \citep{Fabian_Barcons1992}.   This  led  to
suggestions that  Seyfert-2 AGN, which  were observed to  have flatter
X-ray spectra compared  to UV bright QSOs, could  potentially be major
contributors   to   the   XRB  \citep{Setti_Woltjer1989,   Awaki1991}.
Progress  in   the  interpretation  of   the  X-ray  spectra   of  AGN
\citep[e.g][]{Williams1992, Nandra_Pounds1992, Turner1993, Nandra1993,
  Nandra_Pounds1994}  coupled  with   improved  constraints  on  their
cosmological  evolution \citep{Stocke1991,  Maccacaro1991, Shanks1991,
  Boyle1993} led  to the  seminal work of  \citet{Comastri1995}.  They
demonstrated   that  AGN  under   the  unification   paradigm  \citep{
  Antonucci1993}  could  reproduce  the   XRB  spectrum  up  to  about
100\,keV.  A  key prediction of that  model was the  high incidence of
obscured accreting  supermassive black holes (SMBHs)  in the Universe,
including a fraction ($\approx  25$\% of total population) of luminous
Compton-thick ones ($\rm N_H>10^{24}\,cm^{-2}$).

Observations  with  the \xmm  and  \chandra  confirmed  some of  these
predictions,  lending   support  to  the  basic   assumptions  of  the
\citet{Comastri1995}  model.   Deep   surveys  carried  out  by  those
telescopes resolved almost the entire XRB intensity into point sources
up  to  energies of  few  keV \citep[e.g.][]{Worsley2005}.   Moreover,
follow-up observations  suggested that  the majority of  those sources
are  AGN \citep[e.g.][]{Bauer2004}.  Analysis  of their  X-ray spectra
also demonstrated  that in most of  them the direct  X-ray emission is
reprocessed  by  gas  and  dust  clouds  with  column  densities  $\rm
N_H\approx10^{22}-10^{24}\,cm^{-2}$   \citep{Tozzi2006,   Akylas2006}.
X-ray surveys by those telescopes revealed an unexpected dependence of
the  obscured ($\rm N_H>10^{22}\,cm^{-2}$)  AGN fraction  on accretion
luminosity,  which  implies  deviations  from  the  basic  unification
paradigm         \citep[][but         see         \citealt{Dwelly2005,
    DwellyPage2006}]{Ueda2003,        LaFranca2005,        Akylas2006,
  Hasinger2008}.

In  the  wake of  those  developments,  \citet{Gilli2007} revised  the
\citet{Comastri1995}  XRB synthesis model  by including  the knowledge
accumulated  by  the  \xmm  and  \chandra  observations  both  on  the
evolution of the  X-ray luminosity function and the  dependence of the
obscured AGN  fraction on luminosity.  Large  numbers of Compton-thick
sources with column densities $\rm 10^{24-26}cm^{-2}$ (equal number to
that of Compton-thin AGN) were required by the revised model.  Most of
them were predicted to lie  close to, or below, the sensitivity limits
of  the deepest  current  X-ray surveys.   Attempts  to identify  this
hidden population either  directly at X-rays \citep[e.g.][]{Tozzi2006,
  Georgantopoulos2009, Comastri2011, Feruglio2011} or the mid-infrared
\citep[e.g.][]{Fiore2009} suggested  agreement, at least  to the first
approximation,  with the \citet{Gilli2007}  predictions.  At  the same
time, however,  criticism has been  raised on the large  percentage of
Compton-thick AGN adopted by  that model.  Studies selecting local AGN
either      at     hard      X-ray     energies      $\ga     15$\,keV
\citep[e.g.][]{Beckmann2009, Tueller2011,  Burlon2011}, or via optical
spectroscopy   \citep[e.g.][]{Akylas2009}  or   at   the  mid-infrared
\citep[e.g.][]{BrightmanNandra2011_model}  determined  {\it intrinsic}
Compton-thick fractions in the range 15-25\% of the AGN population.

It has also been pointed  out that observational uncertainties as well
as aliases among  the large number of parameters  that are unavoidably
adopted  in  XRB  synthesis  codes,   make  it  hard  to  draw  robust
conclusions on the space density of Compton-thick AGN in the Universe.
\cite{Treister2009},  for example,  showed that  the current  level of
uncertainty  in the  normalisation  of the  XRB  intensity limits  the
predictive  power   of  XRB  synthesis   codes  on  the   fraction  of
Compton-thick AGN.  Moreover, certain  AGN spectral components are not
yet well understood,  e.g the strength of the  radiation reflected off
the torus  or the accretion  disk in the  line of sight.   Tuning this
parameter  in  XRB  synthesis  models  to values  allowed  by  current
observational   constraints  can   change  the   required   number  of
Compton-thick  sources  \citep{Treister2009}.   \cite{DraperBalla2009}
also highlighted  the significance  of unobscured AGN  populations for
the interpretation  of the hard ($>10$\,keV)  X-ray background.  These
authors estimated  the contribution to  the XRB of blazars  with X-ray
spectra dominated  by synchrotron radiation.   Including those sources
can  reduce  the  fraction  of  Compton-thick  QSOs  required  by  the
\cite{Gilli2007} model by about 20\%.

In light of  those studies, it is interesting  to revisit the question
of  whether  the XRB  hump  at  20-30\,keV  requires the  presence  of
Compton-thick  AGN.  This  paper  takes a  critical  approach to  this
problem by presenting an XRB synthesis code that uses state-of-the-art
numerical  simulations  to  model   the  high-energy  spectra  of  AGN
\citep{BrightmanNandra2011_model}.  A feature  of those simulations is
the  realistic modelling  of the  Compton scattering  of photons  in a
dense medium, which has a strong impact on the X-ray spectra of AGN at
obscurations close  to and above  the Compton-thick limit, $\rm  N_H =
10^{24}\,cm^{-2}$.  We then identify  four input parameters to the XRB
synthesis  code (one  of which  is  the number  of Compton-thick  AGN,
others relate to the  AGN spectral characteristics), which encapsulate
the minimum  level of uncertainty in reconstructing  the XRB spectrum.
Instead of fixing those parameters to literature values, we allow them
to vary  within plausible  intervals.  We then  map the volume  of the
space   allowed   to  those   parameters   by  current   observational
determinations  of the  XRB  intensity.  This  approach  allows us  to
investigate in a quantitative  way degeneracies among the AGN spectral
parameters and their impact  on the determination of the Compton-thick
AGN  fraction from  measurements  of the  XRB  spectrum.  Finally,  we
demonstrate  how current data  and future  X-ray missions  can provide
significant  progress   in  our   understanding  of  the   origin  and
composition  of   XRB,  by  directly  constraining   the  fraction  of
Compton-thick AGN  across cosmic time.   In our calculations  we adopt
$\rm H_{0} =  70 \, km \, s^{-1} \, Mpc^{-1}$,  $\rm \Omega_{M} = 0.3$
and $\rm \Omega_{\Lambda} = 0.7$.

\section{The X-ray background model}

The synthesis of  the XRB spectrum requires assumptions  to be made on
(i) the  shape of the spectral  energy distribution of  AGN at X-rays,
(ii)  the volume  density of  AGN  in the  Universe as  a function  of
accretion luminosity  and redshift, (iii)  the distribution of  AGN in
obscuration,  which  is often  parametrised  by  the hydrogen  column
density,  $\rm N_H$.   Each component  of the  XRB synthesis  model is
described below.

\subsection{AGN X-ray spectral energy distribution}

Active  Galactic Nuclei  have complex  spectra at  high  energies that
depend on the  properties of the SMBH (mass, spin),  the rate at which
matter accretes onto it, and on the overall structure and distribution
of  material   in  its   vicinity  (e.g.  torus).    Observations  and
theoretical work suggest that the components needed to reconstruct the
X-ray  spectra of  AGN include  (i)  a power-law  with an  exponential
cutoff at high energies, which represents the intrinsic AGN radiation,
(ii) photoelectric  absorption and Compton  scattering, (iii) emission
lines, the  most prominent  of which is  the FeK$\alpha$  at 6.4\,keV,
(iv)  reflection of  the  direct  AGN radiation  on  an optical  thick
medium,  (v) an excess  emission over  the power-law  extrapolation at
soft  energies ($\rm <1-2  \,keV$), and  (vi) a  possible contribution
from radio jets.

The intrinsic AGN  X-ray spectrum can be parameterised  by a power-law
with photon index $\Gamma$ and an exponential decline at high energies
with e-folding parameter $E_C$. This  functional form is the result of
comptonisation  processes.   Low-energy  photons (UV,  soft  X-rays),
possibly associated with the  accretion disk, are upscattered to hard
X-ray  energies  by the  energetic  particles  of  a hot  corona  that
surrounds       the       innermost       regions       of       SMBHs
\citep[e.g.][]{Haardt1991}. In  this standard model,  the photon index
$\Gamma$ is related to the  temperature, the optical depth and overall
structure   of    the   matter    close   to   the    central   engine
\citep[e.g.][]{Murphy2011}.   The   energy  cut-off  depends   on  the
temperature of the thermal hot  electrons that are responsible for the
Compton-scattering of the accretion disk seed photons.

The direct X-ray  emission of AGN is modified by  cold material in the
vicinity  of the  SMBH, such  as the  putative torus  of gas  and dust
clouds. X-ray photons that intersect  such clouds are absorbed with an
efficiency that depends on the density and the chemical composition of
obscuring  screen. When  the column  density of  the  absorbing screen
increases, Compton-scattering of X-ray photons on the electrons of the
intervening  material  becomes   increasingly  important  relative  to
photoelectric absorption.  The net result is that the direct emission,
at  least  below  about  10  keV, is  significantly  suppressed.   The
Compton-scattering  optical  depth is  defined  $\tau \simeq  \sigma_T
N_H$,  where $\sigma_T$  is the  Thomson-scattering cross  section and
$N_H$ the hydrogen column density of the material.  Compton-scattering
dominates the fate of the  X-ray photons emitted by the central engine
at     $\tau    \ga     1$    or     equivalently    $\rm     N_H    =
{\sigma_T}^{-1}=1.5\times10^{24}    cm^{-2}$.      The    impact    of
photoelectric absorption  and Compton-scattering on  the intrinsic AGN
spectrum is accounted for in a self-consistent way using the numerical
simulations of \cite{BrightmanNandra2011_model}.   We used their model
that  describes the  transmission  spectrum from  an isotropic  source
located  at  the  centre   of  a  uniform  spherical  distribution  of
matter. Input parameters of the simulations are the $\Gamma$ and $E_C$
of the intrinsic AGN spectrum, the column density $\rm N_H$, and metal
abundance  of  the  obscuring   material.   For  simplicity  we  fixed
abundances to solar values.

Observations also show the presence of a prominent iron $K\alpha$ line
in    the   spectra    of    AGN   at    an    energy   of    6.4\,keV
\citep{Nandra_Pounds1994}.   This  can   be  produced  as  direct  AGN
emission intersects  optically thick  material, such as  the accretion
disk  or the  torus, and  is  Compton-backscattered into  the line  of
sight.  The  same processes produces a  reflection continuum component
\citep{GeorgeFabian1991,   Murphy2011},  which   becomes  increasingly
important  at  energies  $>8$   keV.   The  numerical  simulations  of
\cite{BrightmanNandra2011_model}  include  the  $K\alpha$ emission  of
iron as well as the transitions from several other elements.  There is
no  reflection   component  in   the  spherical  symmetric   model  of
\cite{BrightmanNandra2011_model}.  We chose to describe this component
using the  simulations of  \cite{Magdziarz1995} as implemented  in the
{\sc  pexrav} model of  {\sc xspec}.   We parameterise  the reflection
emission  by  its  strength  $f_{REF}$  relative  to  the  direct  AGN
radiation  integrated in  the rest-frame  2--10\,keV band.   Other XRB
synthesis  models \citep[e.g.][]{Gilli2007, Treister2009}  measure the
amplitude of  the reflection component  with the parameter  $R$, which
denotes  the solid angle  subtended by  the reflecting  material.  For
comparison $R$=0.5 \citep{Treister2009} corresponds to $f_{REF}=2$.

 Ideally, the  modelling of  AGN X-ray spectra  should also  include a
 soft-excess  component.  For  type-1  AGN  this is  suggested  to  be
 relativistically   blurred  photoionized  disc   reflection  emission
 \citep[e.g.][]{Crummy2006}.   In  type-2  AGN,  the  soft  excess  is
 proposed to be  scattered radiation in the line of  sight that is not
 absorbed by the  torus \citep[e.g.][]{Turner1997}.  When synthesising
 the  XRB spectrum,  we  chose  not to  include  the above  components
 because  we are  primarily interested  in  the intensity  of the  XRB
 spectrum at  $\ga 2$\,keV.   The contribution of  the soft  excess is
 likely  to  be small  at  those  energies.   These components  become
 important,  however,  when  using  the  XRB synthesis  code  to  make
 predictions  on  the expected  number  of  AGN, particularly  heavily
 obscured  ones,  as a  function  of  X-ray  flux at  relatively  soft
 energies  (e.g.  0.5-2\,keV).   For  this application  we included  a
 soft-excess  scattered-light  component  in  the  spectra  of  type-2
 AGN. It follows the same  functional form as the direct AGN radiation
 (power-law plus  high-energy cutoff) with a normalisation  fixed to 3
 per cent of  the intrinsic AGN spectrum in  the 2-10\,keV band.  This
 is  the same as  the strength  of the  soft-excess component  for the
 type-2 AGN adopted by \cite{Gilli2007}.

About  10\%  of  the  AGN  have  powerful radio  jets  that  may  also
contribute  to or  even dominate  the emission  at  X-ray wavelengths.
Although we  did not  directly include such  a spectral  component, we
used  the  results  of   \cite{DraperBalla2009}  to  account  for  the
contribution of blazars to the XRB spectrum.

Given our assumptions for the main emission components of AGN, the key
parameters for  modelling their intrinsic X-ray  spectra are $\Gamma$,
$E_C$, and the strength  of the reflection component, $f_{REF}$. X-ray
spectroscopy   of  individual  AGN   provides  constraints   on  those
parameters.   There  are, however,  degeneracies  among the  different
emission   components   which   when   combined   with   observational
limitations,  do  not allow  a  description  of  the typical  spectral
characteristics  of  AGN  at  the  accuracy  level  required  for  the
synthesis of the XRB.  For example, the average photon index of AGN is
determined  to be  in the  range  1.8--2.0 with  a standard  deviation
around   the  mean   of   about  0.15-0.30   \citep{Nandra_Pounds1994,
  Dadina2008,   Beckmann2009,   BrightmanNandra2011_model,  Ricci2011,
  Burlon2011}.  Observational  constraints on the  cutoff energy range
from   as   low  as   50\,keV   \citep[e.g.][]{Molina2009}  to   about
$\sim$300\,keV  \citep{Dadina2008}.  The  strength  of the  reflection
component  also  varies among  different  studies, with  observational
constraints  ranging  from  $f_{REF}=0.01$  to  0.1  \citep{Ricci2011}
(corresponding  to a reflection  parameter $R\sim0.2-2$).   A possible
dependence  of  the reflection  of  the  source characteristics,  i.e.
type-1  vs  type-2   AGN,  may  explain  the  wide   range  of  values
\citep{Ricci2011}, although such a trend remains to be established.

\subsection{The AGN X-ray luminosity function}

For the X-ray luminosity function (XLF) of AGNs and its evolution with
redshift  we  used  the  parameterisation of  \citet{Ueda2003}.   They
estimated the AGN XLF in the rest-frame 2-10\,keV energy range using a
combination of  hard-band ($>2$\,keV) surveys conducted  with the {\it
  HEAO-1}, {\it  ASCA}, and {\it Chandra} missions.   Here, we adopted
the luminosity-dependent density evolution of the luminosity function,
which provides  a better  fit to the  \citet{Ueda2003} data.   In this
parameterisation the cutoff redshift  after which the evolution of AGN
stops   increases   with  luminosity.    We   preferred   to  use   the
\citet{Ueda2003}     XLF    over     more     recent    determinations
\citep[e.g][]{Aird2010} because  the former include  a self-consistent
estimation of the  fraction of Compton-thin AGN ($\rm  N_H = 10^{22} -
10^{24} \, cm^{-2}$).

\subsection{The distribution of AGN in hydrogen column density}

It  is assumed  that AGN  are  distributed uniformly  in the  hydrogen
column density logarithmic intervals, $\log N_{H} = 20 - 22$ and $22 -
24$  ($\rm  cm^{-2}$).   The  AGN  $N_H$  distribution,  $f(N_H)$,  is
therefore approximated by  a step function with a  fixed ratio, ${\cal
  R}$, between absorbed ($ N_H=10^{22}  - 10^{24} \rm \, cm^{-2}$) and
unabsorbed ($N_H  < \rm 10^{22}  \, cm^{-2}$) sources.   The parameter
${\cal  R}$ is  a function  of  X-ray luminosity  as parameterised  by
\cite{Akylas2006}.  It increases from  ${\cal R}\approx0.4$ at $L_X (
\rm 2 -  10 \, keV ) \approx 10^{45}$\,\lunits\,  to ${\cal R} \approx
0.8$ at  $L_X ( \rm  2 - 10  \, keV ) \approx  10^{42}$\,\lunits.  For
comparison,  the corresponding  fractions in  the Gilli  et  al. (2007)
model   are  ${\cal   R}  \approx   0.45$  and   0.75  at   the  above
luminosities.   The  adopted   parametrisation   for  the   luminosity
dependence of  the obscured  AGN fraction is  also similar to  that of
\cite{Ueda2003}.   We  chose  to  use the  \cite{Akylas2006}  results
because they  are using X-ray spectral analysis,  not hardness ratios,
to determine the column density of individual AGN in their sample.  It
should be emphasised that our results and conclusions would not change
if  we  used  the   \cite{Ueda2003}  functional  form  for  the  $L_X$
dependence of ${\cal R}$.

Compton-thick  AGN are  not included  in the  \cite{Ueda2003}  XLF, or
indeed any XLF determination  based on observations in the 0.5-10\,keV
spectral band. This is because for $N_H\ga 10^{24}\, \rm cm^{-2}$, the
bulk  of   the  direct  AGN  emission  at   rest-frame  energies  $\la
10-20$\,keV  is  suppressed.    These  heavily  obscured  sources  are
therefore too faint in the 0.5-10\,keV spectral window of \chandra and
{\it XMM-Newton}.  Their identification  is therefore hard even in the
deepest   current   X-ray   survey  fields   \citep[e.g.][]{Tozzi2006,
  Georgantopoulos2009}.   As a result,  the fraction  of Compton-thick
AGN  remains  uncertain (see  Comastri  et  al.   2011 and  references
therein).  We  introduced the parameter $f_{CT}$, which  is defined as
the  fraction of  Compton-thick AGN  ($N_H >  10^{24}\,  \rm cm^{-2}$)
relative  to  mildly  obscured  Compton-thin sources  ($N_H=10^{22}  -
10^{24}\, \rm cm^{-2}$). We  then added this fraction of Compton-thick
AGN  to the  XLF of  \cite{Ueda2003}.   It is  therefore assumed  that
Compton-thick sources follow the same XLF as mildly obscucred systems.

\begin{figure}
\begin{center}
\includegraphics[height=0.9\columnwidth]{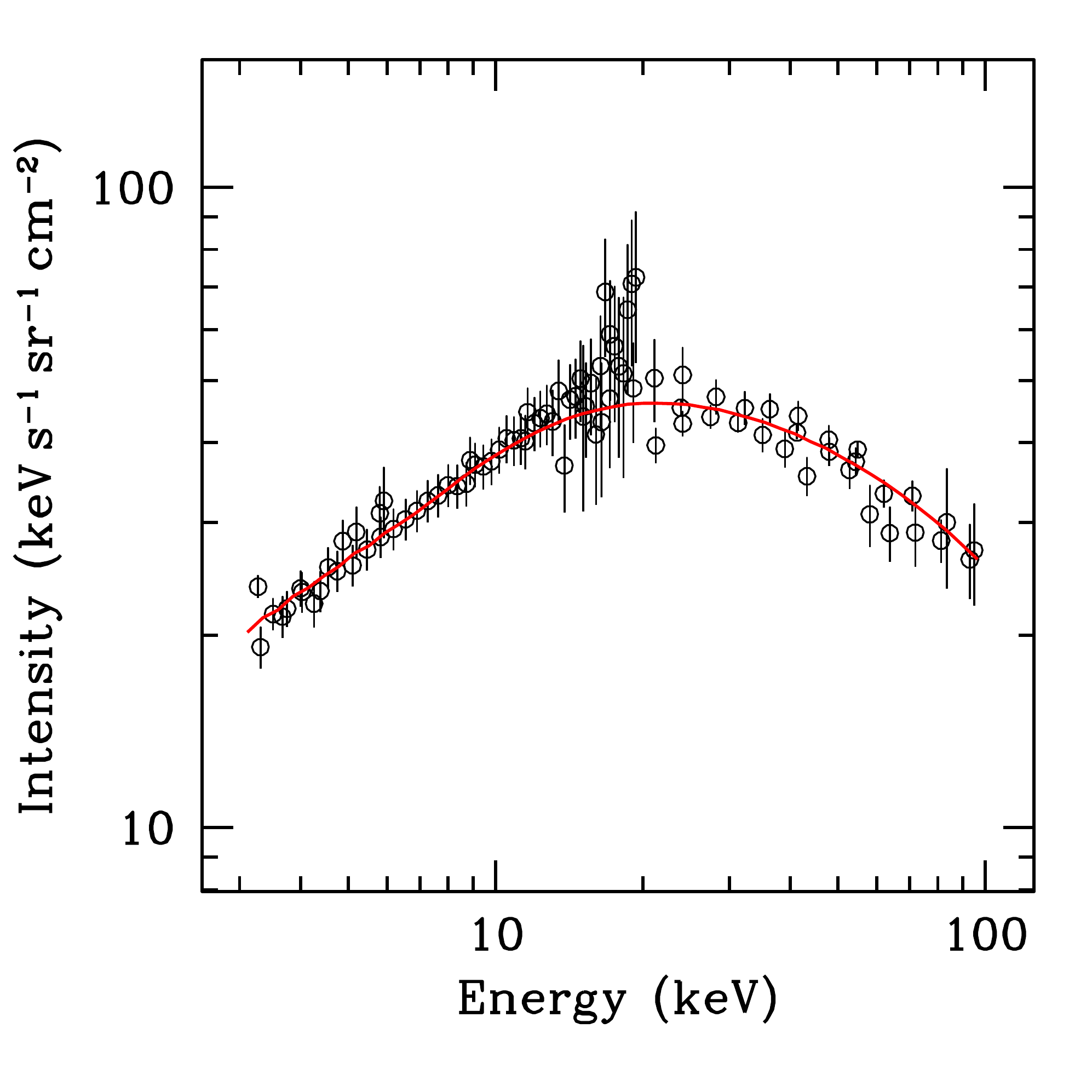}
\end{center}
\caption{Red curve: an example of an XRB spectrum produced by
  our  synthesis model  for parameters  $\Gamma=1.88$, $E_C=230$\,keV,
  $f_{REF}=0.055$ and $f_{CT}=0.1$.   The points are the observational
  constraints on the  intensity of the XRB at  different energies from
  the studies  discussed in the  text.  Comparison of the  model curve
  with the  data yields a $\chi^2$  of 70.8 for 86  degrees of freedom
  (90 data points minus four free parameters).}\label{fig_xrb}
\end{figure}

\begin{figure*}
\begin{center}
\includegraphics[height=0.9\columnwidth]{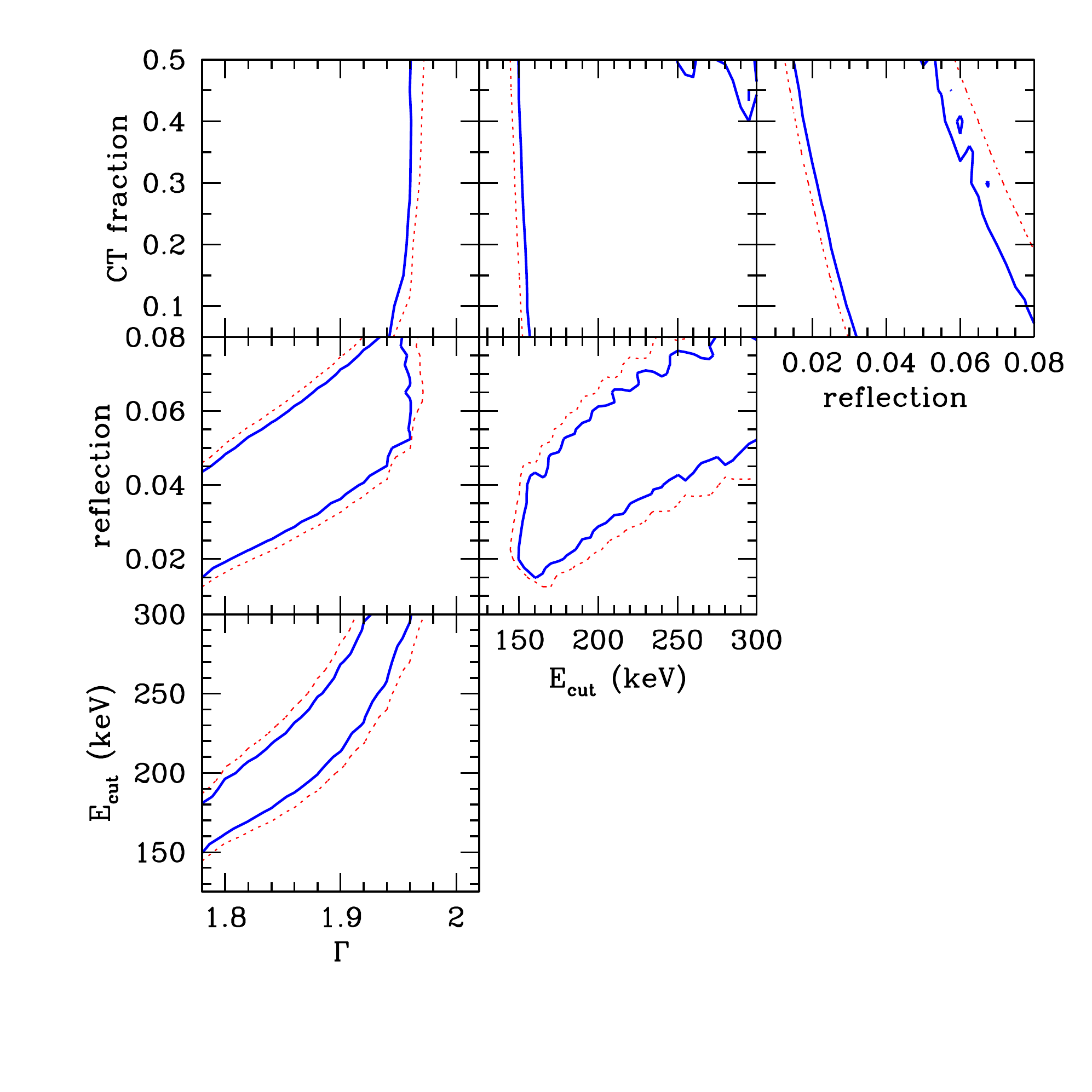}
\includegraphics[height=0.9\columnwidth]{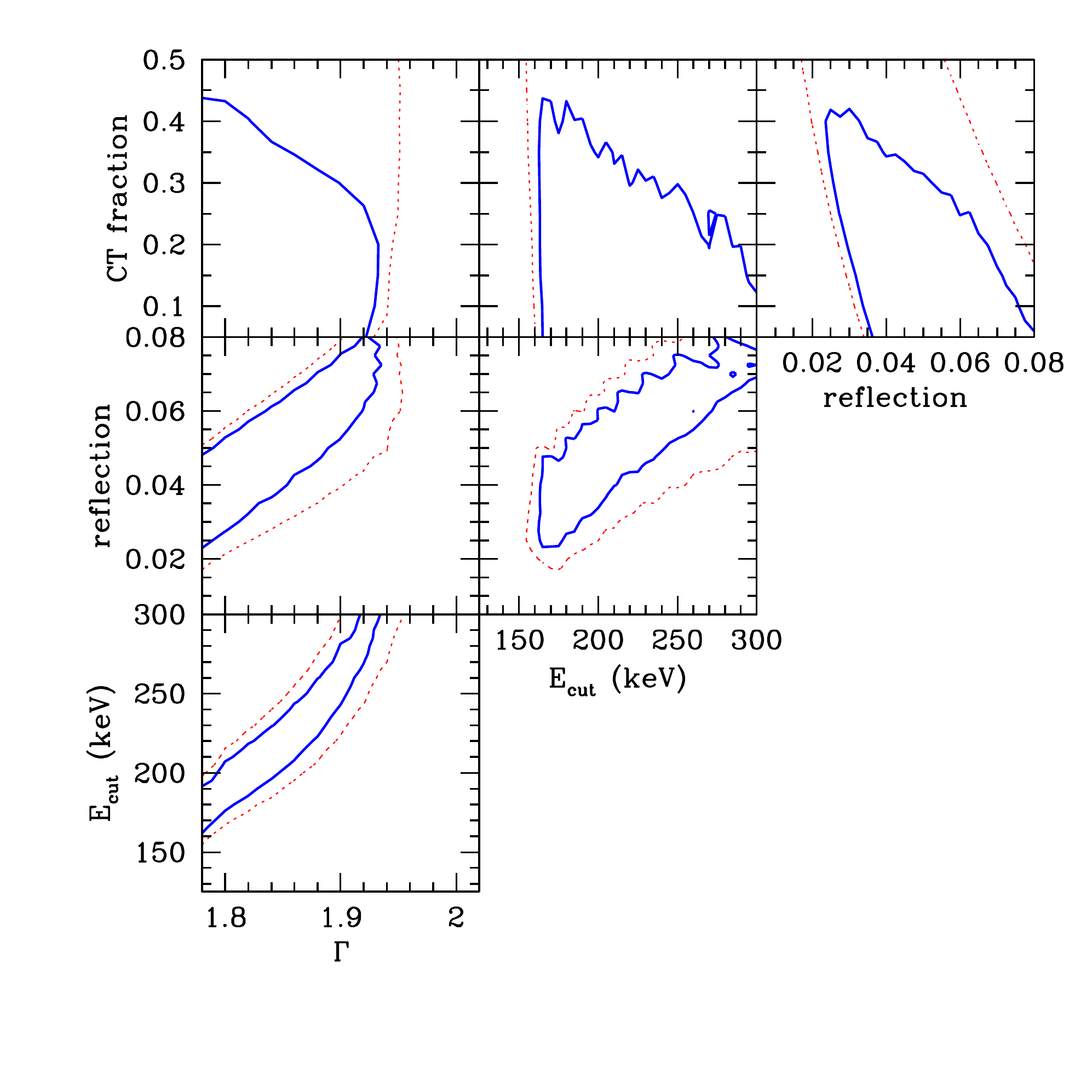}
\end{center}
\caption{{\bf  Left:} confidence  intervals for  the  model parameters
  $\Gamma$, $E_C$,  $f_{REF}$ and $f_{CT}$ used to  synthesise the XRB
  spectrum. The (blue) thick continuous line corresponds to the 68 per
  cent  confidence level ($\Delta  \chi^2=4.72$ for  four parameters).
  The  (red)  thin dotted  line  marks  the  95th percentile  ($\Delta
  \chi^2=9.70$ for four parameters).  {\bf Right:} Same as right panel
  but  including to  our  XRB synthesis  results  the contribution  of
  blazars,      as      modelled      by      \citet{DraperBalla2009}.
}\label{fig_contours}
\end{figure*}

\begin{figure*}
\begin{center}
\includegraphics[height=0.9\columnwidth]{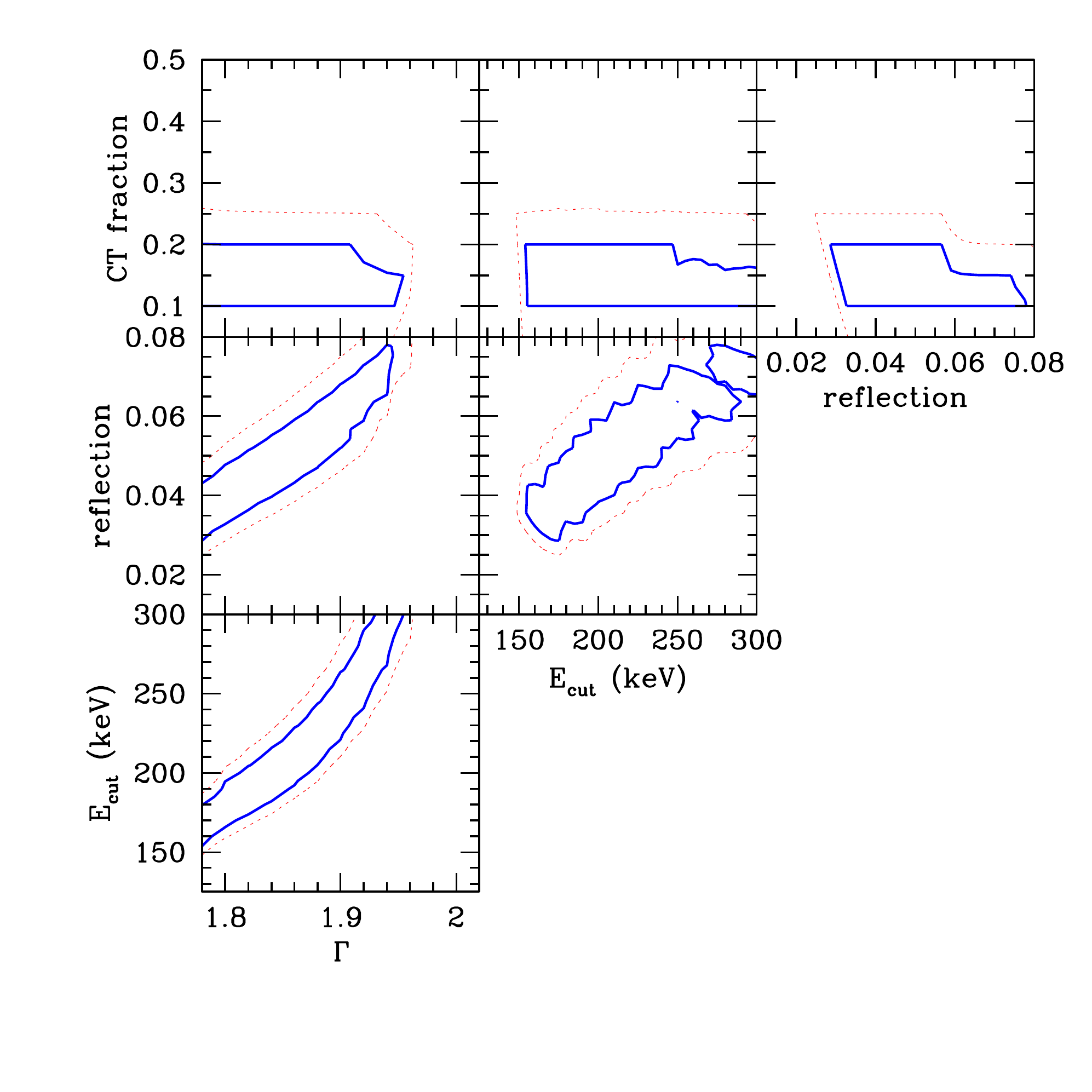}
\includegraphics[height=0.9\columnwidth]{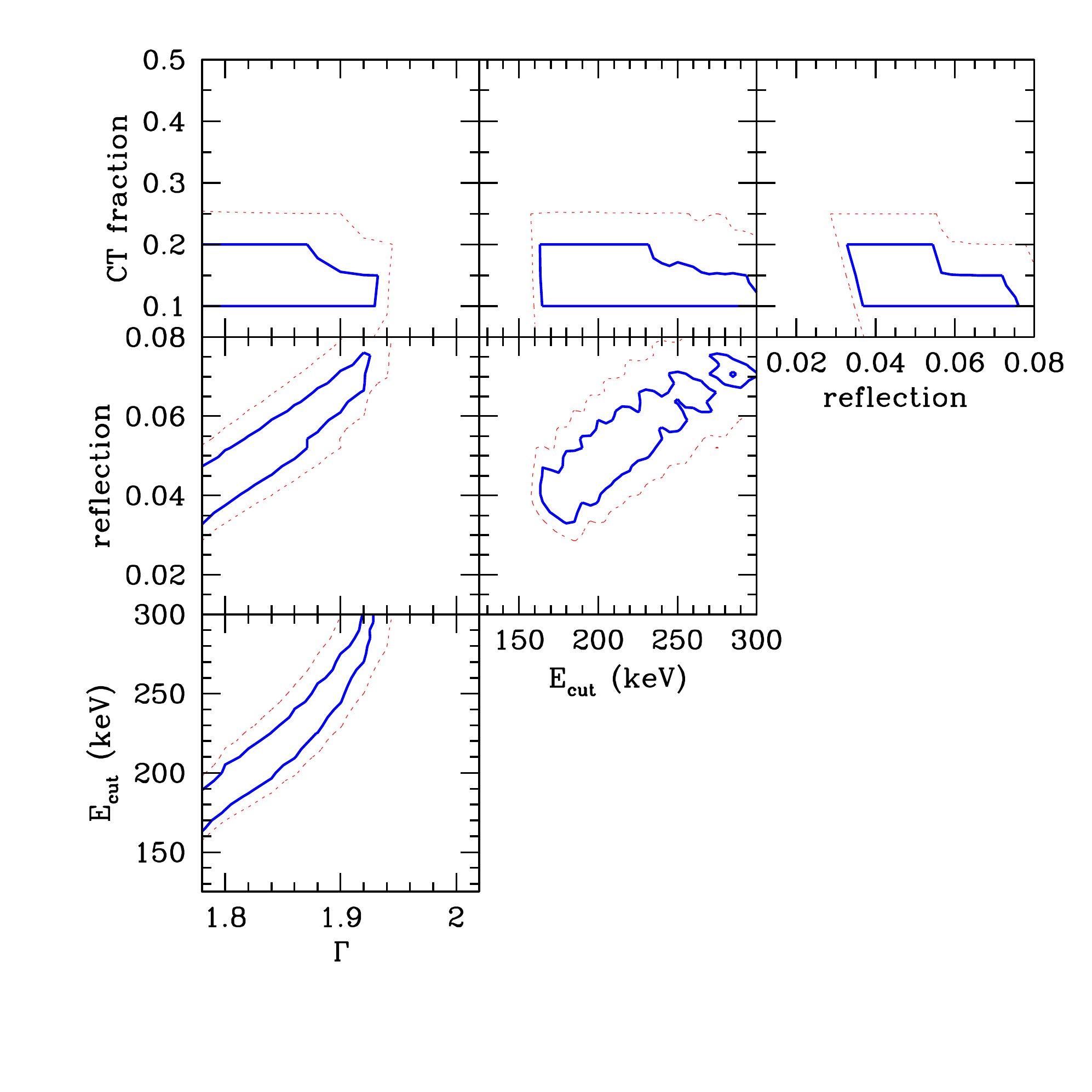}
\end{center}
\caption{Joint   constraints  on   the  parameters   $\Gamma$,  $E_C$,
$f_{REF}$ and $f_{CT}$ based on both the XRB spectrum and the fraction
of   Compton-thick  of  AGN   in  the   {\it  SWIFT/BAT}   AGN  survey
\citep{Burlon2011}.   {\bf  Left:} The  (blue)  thick continuous  line
marks the  region of  the parameter space  which is consistent  at the
68\%  confidence level  with both  the observed  XRB spectrum  and the
fraction of {\it SWIFT/BAT}  Compton-Thick AGN.  The (red) thin dotted
line marks the  region which is consistent with  those observations at
the 95\% confidence level.  {\bf  Right:} Same as left panel including
the contribution of blazars, as modeled by \citet{DraperBalla2009}, to
the XRB synthesis results.}
\label{fig_contours_bat}
\end{figure*}

\begin{table} 
\centering
\caption{Range  of  parameters  used   in modelling  the  XRB
}\label{tab_grid}
\begin{tabular}{cccc} \hline\hline parameter &  min & max & step\\ (1)
& (2) & (3) & (4) \\ 
\hline 

$\Gamma$ & 1.78 & 2.02 & 0.02 \\ 

$E_C$ & 125\,keV & 300\,keV & 5\,keV\\ 

$f_{REF}$ & 0.01 & 0.08 & 0.0025 \\ 

$f_{CT}$ & 0.05 & 0.50 & 0.05 \\ \hline

\end{tabular}
\begin{list}{}{}
\item{1:  parameter of  interest,  where $\Gamma$  and  $E_C$ are  the
  power-law index and high-energy  cutoff of the intrinsic AGN
  spectrum,  $f_{REF}$  is   the  reflection  component  strength  and
  $f_{CT}$ is  the fraction of  Compton-thick sources relative  to the
  obscured AGN  population; 2: minimum  value of parameter;  3: maximum
  value of parameter; 4: step size}
\end{list}
\end{table}

\section{Adopted parameters}

Modelling  the  XRB  involves  many  parameters,  some  of  which  are
reasonably  well-constrained (e.g.   the XLF  of Compton  thin  AGN at
least out  to $z=2-3$), while  others have larger  uncertainties (e.g.
AGN X-ray spectrum). We  identified four parameters that are important
for  shaping the  XRB  spectrum and  for  which sufficiently  accurate
observational determinations  are still  not available. These  are the
photon  index, $\Gamma$,  and the  high-energy cutoff,  $E_c$,  of the
intrinsic AGN  spectrum, the fraction  of the reflection  component in
AGN  X-ray  spectra,  $f_{REF}$,  and the  fraction  of  Compton-thick
sources, $f_{CT}$,  among AGN. Instead  of fixing those  parameters to
plausible values, we adopted  the alternative approach of leaving them
free  to study  the region  of the  four-dimensional  parameter space,
which  is consistent  with current  observational data.  These include
measurements  of   the  XRB  spectrum  itself  and   the  fraction  of
Compton-thick  AGN   detected  by  high-energy   ($\rm  \ga  10\,keV$)
missions.

We emphasise  that the list  of free parameters  we chose is  far from
exhaustive.   For  example,  a  possible redshift  dependence  of  the
fraction     of      obscured     AGN     \citep[e.g.][]{LaFranca2005,
  TreisterUrry2006,  DellaCeca2008},  which  is  not included  in  our
modelling, would  imprint on  the predicted XRB  spectrum.  Similarly,
the results of the synthesis code depend on the spread of the $\Gamma$
distribution,  which we  fixed to  0.15 following  the  {\it BeppoSAX}
results of \citet{Dadina2008}.  Another limitation of our modelling is
that  we assumed  that  all AGN  have  the same  values  of $E_c$  and
$f_{REF}$,  when a  distribution of  values might  be  more realistic.
Nevertheless,   by   investigating   the   impact   of   observational
uncertainties of  a few  parameters on the  reconstruction of  the XRB
spectrum,  one can assess  whether observations  of the  diffuse X-ray
background intensity can place reliable constraints on AGN properties.
Table \ref{tab_grid} shows the adopted range of values for each of the
four parameters  of interest as well  as the step size  used to sample
the parameter space.

For synthesising the XRB spectrum,  the assumed XLF and $N_H$ function
were integrated  in redshift, luminosity, and  hydrogen column density
to estimate  the expected number  of AGN at  a given bin  of redshift,
luminosity,  and  $N_H$.   The  integration  was carried  out  in  the
luminosity      and     column     density      intervals     $L_X(\rm
2-10\,keV)=10^{40}-10^{47}\,      erg       \,      s^{-1}$,      $\rm
N_H=10^{20}-10^{25}\,  cm^{-2}$.   The   redshift  range  adopted  was
z=0-7. Limiting the  integration to $z=4$, to explore  the impact of a
possible   exponential  decline  in   the  AGN   XLF  at   $z\ga  3-4$
\citep[e.g.][]{Brusa2009, Civano2011}, changes the estimated intensity
of the XRB  spectrum by less than  1\%.  The step in $z$  and $L_X$ is
logarithmic and variable in size, taking a minimum value of $\rm d\log
z  =0.1$  at $z=1.5$  and  $\rm  d\log L_X  =  0.3$  at $\log  L_X(\rm
2-10\,keV)=43.8$ $\rm  erg \,  s^{-1}$ and increasing  at lower/higher
redshift/luminosities.   This  significantly  improves  the  computing
performance of  the code without loss  of accuracy because  the XRB is
dominated by  AGN at $z  \approx 1 -  2$ and $L_X(\rm  2 - 10  \, keV)
\approx 10^{43} - 10^{45}  \, erg \, s^{-1}$ \citep[e.g.][]{Aird2010}.
The $\rm  N_H$ step size  is fixed to  $\rm \Delta \log  N_H=0.5$ $\rm
cm^{-2}$.

At each  $z$, $L_X$ and  $\rm N_H$ bin  the observed shape of  the AGN
spectrum at energies 3-100\,keV was determined using the simulations of
\cite{BrightmanNandra2011_model} to model  the effect of photoelectric
absorption and  Compton-scattering for a given  intrinsic photon index
$\Gamma$  and high-energy  cutoff $E_c$.   A Gaussian  distribution of
photon indices was adopted, with a fixed spread of $\sigma=0.15$ around
the mean.  Reflection was also added to the AGN spectrum using the {\sc
pexrav} model of  {\sc xspec} and assuming a  fixed fraction $f_{REF}$
of this component relative to the direct AGN emission in the 2-10\,keV
band. The resulting  spectrum was then normalised to  the number of AGN
within a  given $z$, $L_X$  and $\rm N_H$  bin as determined  from the
integration of the  XLF.  Adding the normalised spectra  from all $z$,
$L_X$ and  $\rm N_H$ bins  yields the XRB  intensity as a  function of
energy for  a given set  of parameters $\Gamma$, $E_c$,  $f_{REF}$ and
$f_{CT}$. Figure  \ref{fig_xrb} shows an  example of a  model spectrum
produced by our code.

The  standard  $\chi^2$  statistic  was  estimated  for  each  set  of
parameters  ($\Gamma$, $E_c$,  $f_{REF}$, $f_{CT}$)  by  comparing the
model  predictions with  observations at  discrete energies.   In this
exercise we used measurements of the intensity of the XRB estimated by
\citet[][RXTE/PCA,    3-20\,keV]{Revnivtsev2003},   \citet[][INTEGRAL,
  5-100\,keV]{Churazov2007},         \citet[][{\it         SWIFT/BAT},
  15-200\,keV]{Ajello2008_xrb},        \citet[][{\it        SWIFT/XRT}
  1.5-7\,keV]{Moretti2009}         and         \citet[][INTEGRAL/IBIS,
  20-200\,keV]{Turler2010}.  These measurements  are plotted in Figure
\ref{fig_xrb}.  Observational studies on the XRB spectrum that provide
confidence  intervals, not  discrete  data points  (e.g.  {\it  ASCA};
\citealt{Gendreau1995},    \citealt{Kushino2002};    {\it    Chandra},
\citealt{Hickox2006};     {\it     XMM-Newton},    \citealt{Lumb2002},
\citealt{DeLuca_Molendi2004};  {\it  BeppoSAX},  \citealt{Vecchi1999})
were not used  in the analysis.  The 1\,keV  normalisations of the XRB
spectrum determined from those studies are in the range $9.6-12 \rm \,
keV~\,keV^{-1}~\,cm^{-2}~\,s^{-1}~\,sr^{-1}$.  This should be compared
with       the       1\,keV       normalisation      of       $12.2\rm
\,keV~\,keV^{-1}~\,cm^{-2}~\,s^{-1}~\,sr^{-1}$ determined  by the {\it
  SWIFT/XRT}, which was  used as a constraint in  our analysis.  Also,
the    XRB    measurements    from    the   {\it    HEAO-1}    mission
\citep{Marshall1980}  were  not compared  with  the model  predictions
since there  is a  controversy over the  absolute flux  calibration of
that mission, which may result  in a systematic underestimation of the
XRB     intensity    level     \citep[e.g][]{Ueda2003,    Worsley2005,
  Treister2005}.  Indeed, almost all measurements of the XRB intensity
at 1\,keV lie above the {\it HEAO-1} estimate by factors of up to 1.5.
Stray-light  affecting  focusing  X-ray  telescopes, unlike  the  {\it
  HEAO-1}, may also be responsible for the discrepancy.  Nevertheless,
{\it INTEGRAL} and {\it SWIFT/BAT},  which have similar optics to {\it
  HEAO-1},  also found  a higher  normalisation for  the  XRB intensity
\citep{Churazov2007, Moretti2009}.

The comparison between model and observations is limited to the energy
range  3-100\,keV. Below  about  3\,keV sources  other  than AGN  also
contribute to the  XRB, i.e.  X-ray emission from  our Galaxy, groups,
and  clusters  \citep{Blanchard1992}.    At  energies  above  100\,keV
processes that are not modelled  in this paper, such as inverse-Compton
scattering of photons in relativistic jets of radio-loud AGN, dominate
\citep[see][]{DraperBalla2009}. These  processes are also  expected to
have a small contribution to the XRB below 100\,keV \citep{Giommi2006,
  Ajello2009}.   We approximated  their impact  by adding  to  the XRB
spectrum estimated  by our synthesis  code the Blazar  contribution as
modelled by  \citet {DraperBalla2009}, assuming  a duty cycle  of 13\%
for the BL Lacs.  In that  model the blazar contribution to the XRB is
12\% in  the 0.5-2\,keV band,  7\% in the  2-10\,keV band, 9\%  in the
15-55\,keV band  and close to 100\%  in the MeV  region.  The $\chi^2$
statistic  was  re-estimated for  each  set  of  our model  parameters
($\Gamma$, $E_c$, $f_{REF}$,  $f_{CT}$), including the contribution of
blazars.

The XRB synthesis code also produces predictions on the number density
of  AGN   samples  that   fulfill  different  selection   criteria  in
luminosity,  redshift, and/or obscuration.   The observed  fraction of
Compton-thick AGN  as a  function of flux  limit and  detection energy
band provides constraints on the parameters of the XRB synthesis model
that are independent of those obtained from the XRB spectrum.  In this
exercise we  used the  fraction of Compton-thick  AGN detected  in the
15-55\,keV    energy   band    by   the    {\it    SWIFT/BAT}   survey
\citep{Ajello2008_survey}.    The  sky   coverage   sensitivity  curve
corresponding  to  the 5\,$\sigma$  detection  threshold  of the  {\it
  SWIFT/BAT} sample  was folded in  our XRB synthesis code  to predict
the fraction of  Compton-thick sources in that survey  for the grid of
model  parameters presented  in Table  \ref{tab_grid}.  This  was then
compared with the  observations using binomial distribution confidence
intervals.  There are  197 AGN in the {\it  SWIFT/BAT} sample of which
nine (4.5\%) are  classified Compton-thick \citep{Burlon2011}.  The 68
and 95\% binomial confidence  intervals for the Compton-thick fraction
are 0.031--0.060 and 0.016--0.074, respectively

\section{Results}

Figure \ref{fig_contours}  plots the two-dimensional  projections of the
four-dimensional  parameter  space,  which  is  consistent  with  the  XRB
observations.   All possible  pairs of  the parameters  $\Gamma$,
$E_c$,  $f_{REF}$, and  $f_{CT}$ are  shown in  different  panels.  The
contours correspond to  the 68 and 95\% confidence  intervals.  In the
case  of four degrees  of freedom  these correspond  to $\Delta\chi^2$
relative to the minimum $\chi^2$  value of 4.72 and 9.70, respectively.
The  right set  of  panels in  Figure  \ref{fig_contours} include  the
contribution    of    blazars   to    the    XRB    as   modelled    by
\citet{DraperBalla2009}.

The  parameter space  shown in  Figure \ref{fig_contours}  is complex.
Firstly, there are degeneracies among parameters, such as $\Gamma$ and
$E_C$, with  higher values of  the photon index also  requiring higher
cutoff   energies   to   provide    acceptable   fits   to   the   XRB
observations. There are also effects  related to the resolution of the
parameter grid.  These are manifested  by wiggles in the contours e.g.
in the plots  of $f_{CT}$ vs $E_C$ or  $f_{REF}$ vs $E_C$.  Increasing
the sampling rate of the  parameters (i.e.  smaller step size in Table
\ref{tab_grid})  would  smoothen  the   contours  at  the  expense  of
computation time.

Despite this  limitation, the XRB  spectrum does place  constraints on
some of the four parameters plotted in Figure \ref{fig_contours}.  For
example,  there is a  relatively narrow  region of  the $\Gamma$-$E_C$
parameter space that is consistent  with the observations at the 68 or
the 95  \% confidence interval.   In contrast, all plots  that include
the   $f_{CT}$    parameter   (top    row   of   panels    in   Figure
\ref{fig_contours}) have  confidence contours  that cover most  of the
grid space.   Therefore the parameter  that is least sensitive  to the
XRB  spectrum  (and  hence  least  constrained)  is  the  fraction  of
Compton-thick AGN.   For any value of $f_{CT}$  there are combinations
of the  other three parameters that  yield acceptable fits  to the XRB
spectrum at the 68 or 95\% confidence intervals.

Including  blazars  into  the  X-ray background  synthesis  modelling
reduces the  volume of the  parameter space, which is  consistent with
the XRB  spectrum at  the 68\% confidence  level.  This trend  is more
pronounced for  $f_{CT}$.  Blazars have hard  X-ray spectra, dominated
by synchrotron  emission, and  therefore their net  impact on  the XRB
modelling is  the reduction of  the fraction of  Compton-thick sources
needed to  reproduce the  shape and overall  normalisation of  the XRB
spectrum  \citep{DraperBalla2009}.   Nevertheless,  blazars appear  to
have a  small impact on the  95\% confidence level  contours of Figure
\ref{fig_contours},  which  remain wide,  especially  for panels  that
include the $f_{CT}$ parameter.

Figure   \ref{fig_contours_bat}  shows   how  the   ($\Gamma$,  $E_C$,
$f_{REF}$, $f_{CT}$)  parameter space  is modified by  considering the
joint constraints from both the XRB spectrum and the observed fraction
of Compton-thick AGN in the {\it SWIFT/BAT} survey \citep{Burlon2011}.
As expected, the most affected parameter is $f_{CT}$, which is limited
to the range 10-20\% (68  per cent confidence level).  This is similar
to the Compton-thick  AGN fraction in the local  Universe estimated by
optical and IR surveys \citep{Akylas2009, BrightmanNandra2011_model}.

\begin{figure*}

\begin{center}

\includegraphics[height=8cm, width=8cm]{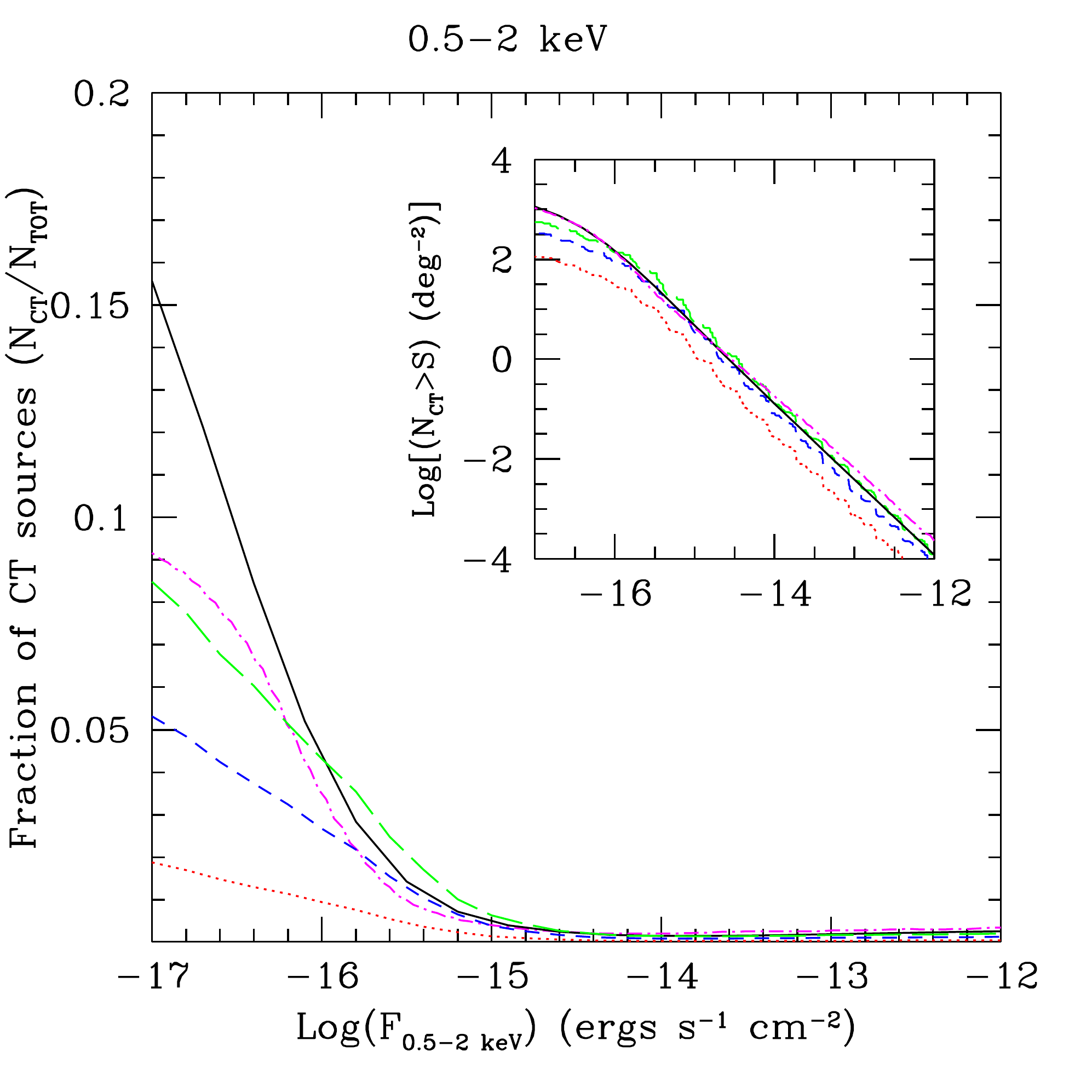}
\includegraphics[height=8cm, width=8cm]{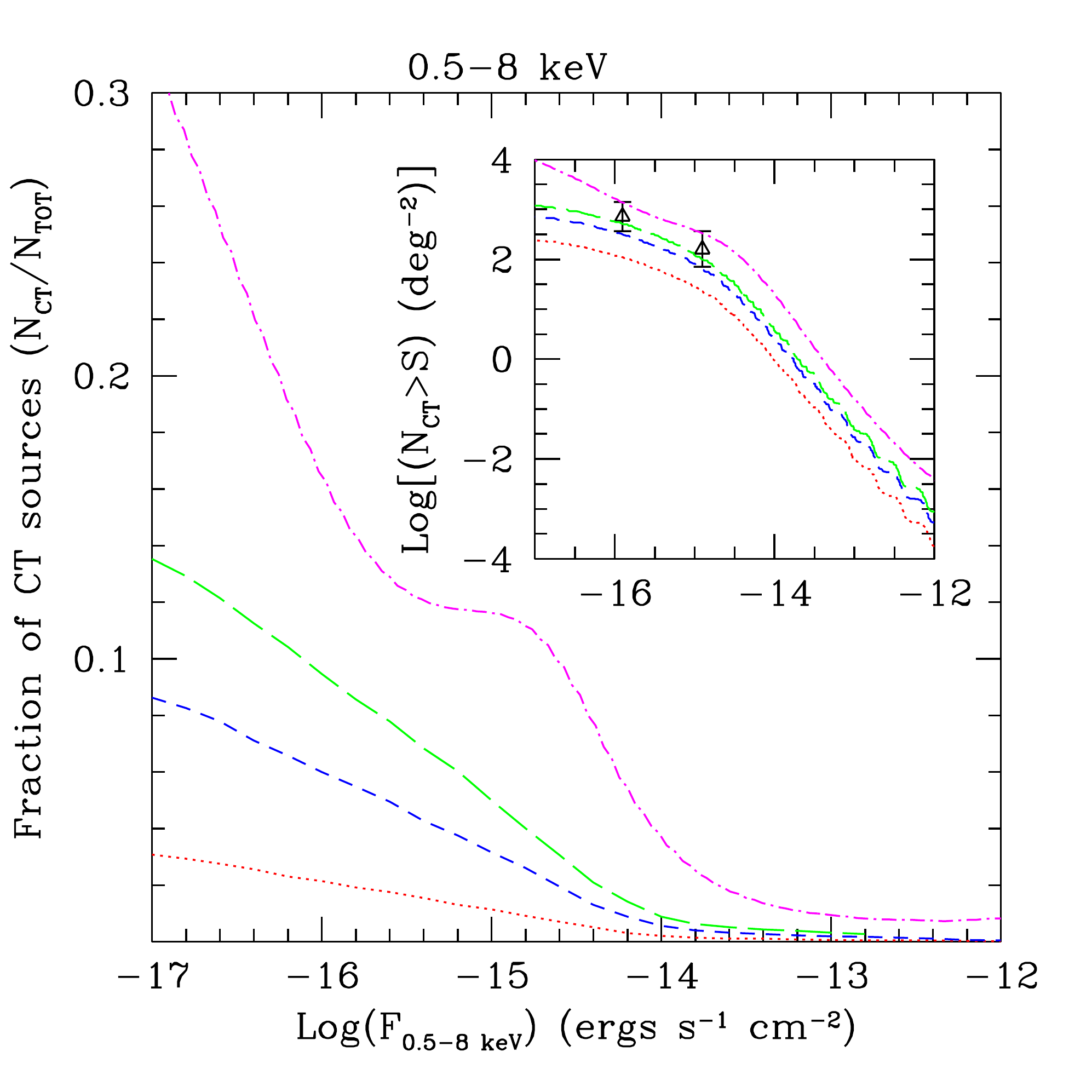}
\includegraphics[height=8cm, width=8cm]{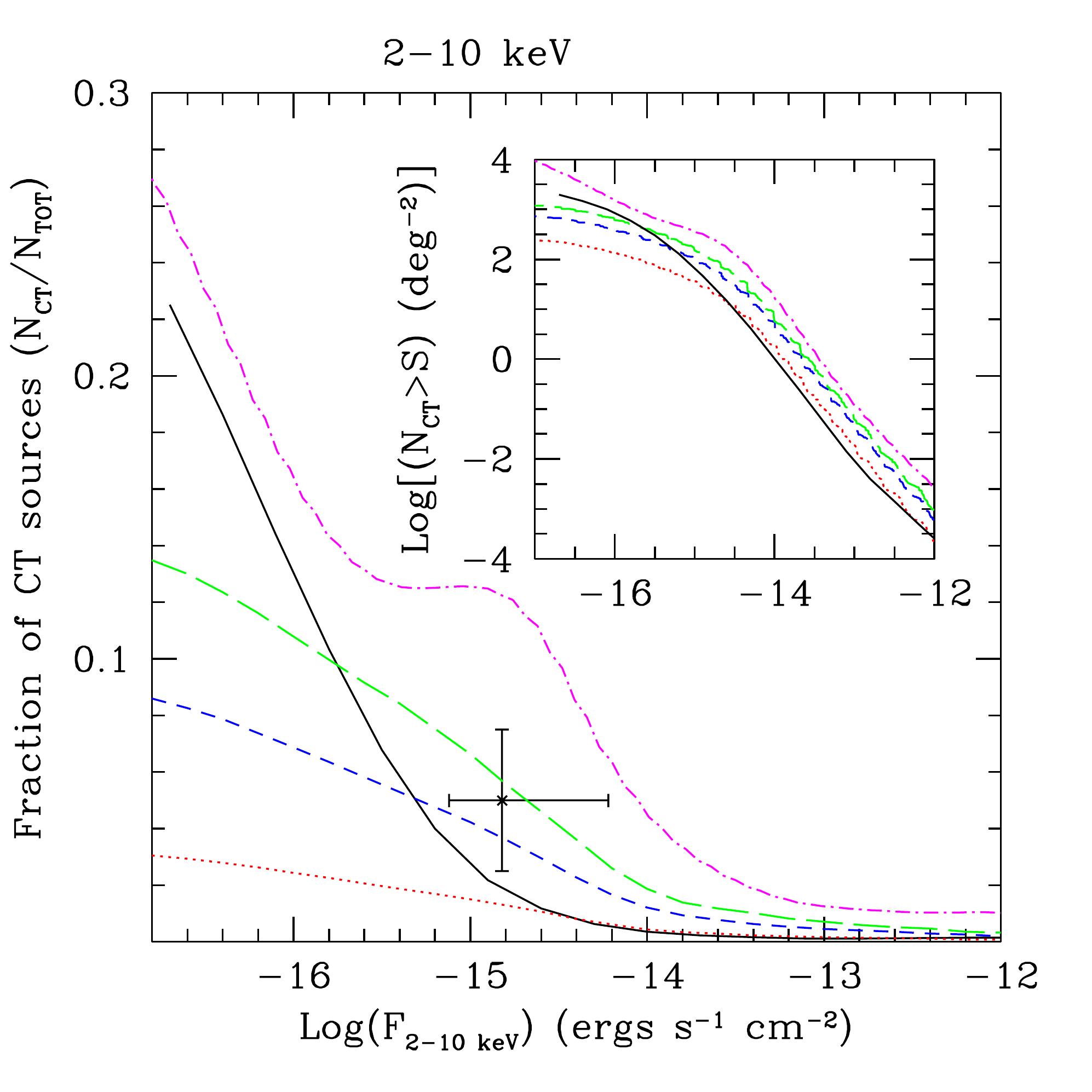}
\includegraphics[height=8cm, width=8cm]{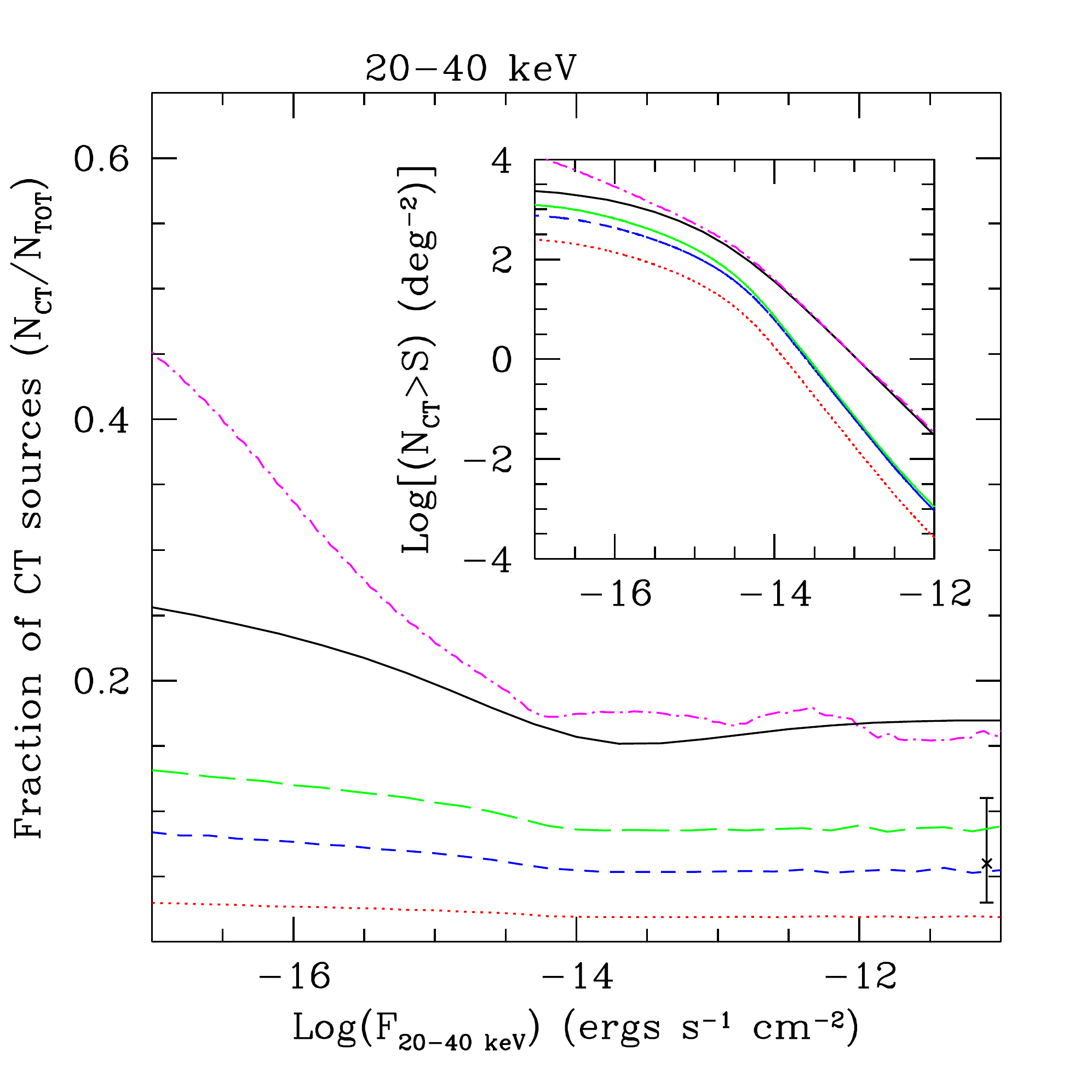}

\caption{The observed  fraction of  Compton-thick AGN in  the spectral
bands 0.5-2, 0.5-8, 2-10 and 20-40\,keV.  Our model predictions for an
intrinsic Compton-thick AGN fraction of  5, 15, 25\% correspond to the
the  red  dotted, blue  short-dashed  and  green  long dashed  curves,
respectively. The black  solid lines are the predictions  of the X-ray
background   synthesis  model   of  \citet{Gilli2007}.    The  magenta
dot-dashed  lines are  the predictions  of  the \citet{Ballantyne2011}
model. The inset plots show  the predictions for the cumulative number
counts of  Compton-thick AGN. The curves  are the same as  in the main
panels.  The observational datapoints are from \citet{Tozzi2006} (2-10
keV),  \citet{BrightmanUeda2012}   (0.5-8  keV),  \citet{Krivonos2007}
adapted from \citet{Treister2009} (20-40keV)}.
\label{ncounts}

\end{center}

\end{figure*}

\begin{figure*}
\begin{center}

\includegraphics[height=8cm, width=8cm]{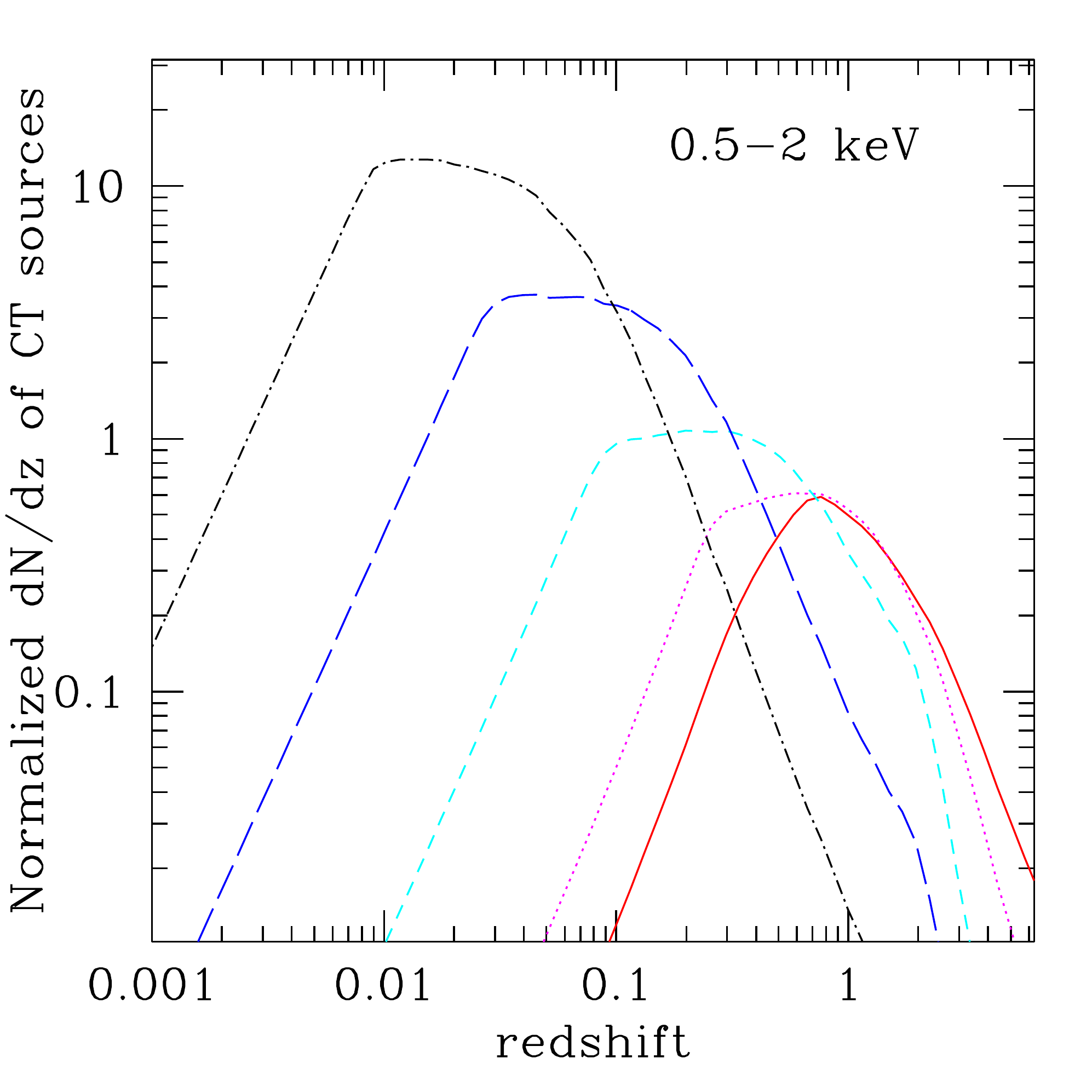}
\includegraphics[height=8cm, width=8cm]{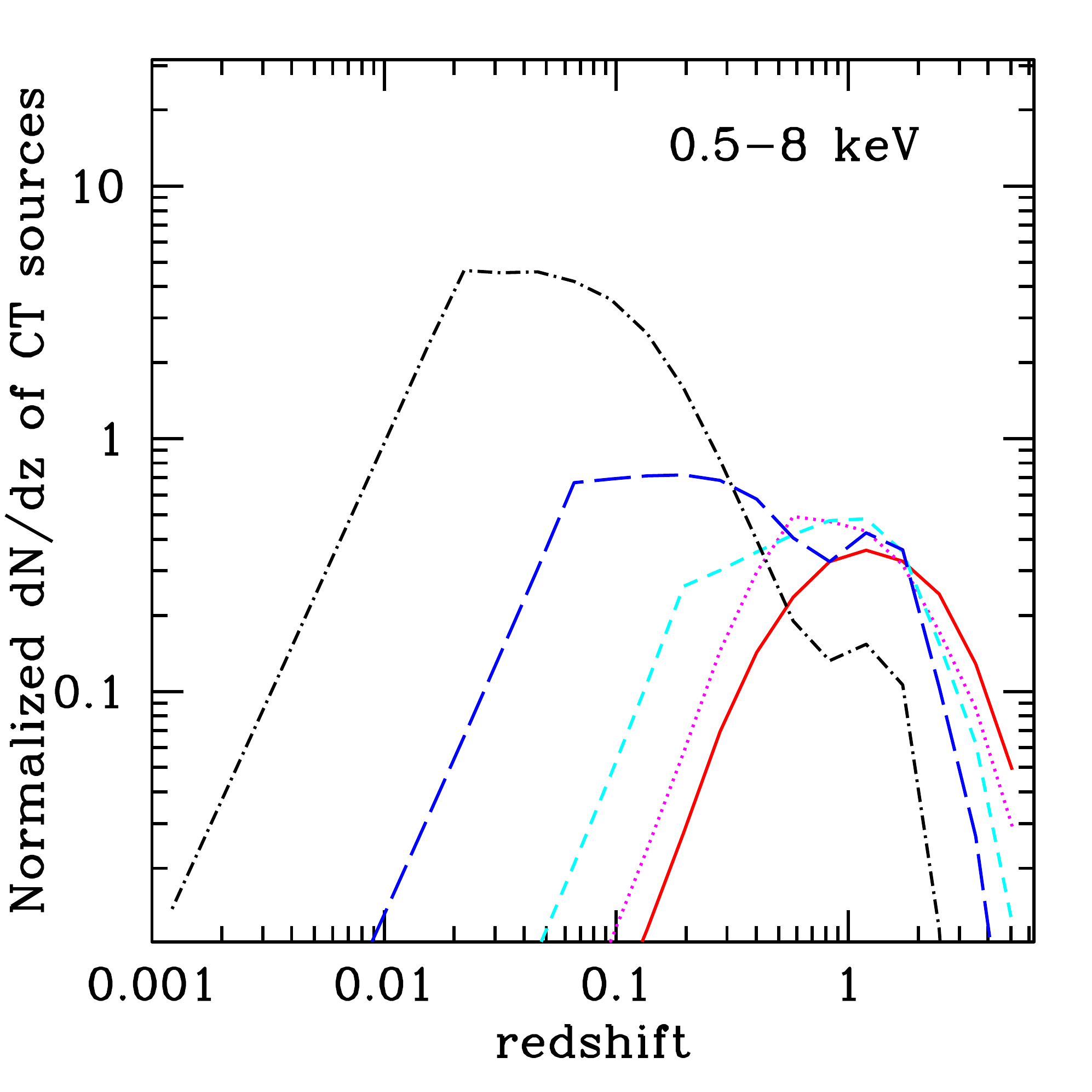}
\includegraphics[height=8cm, width=8cm]{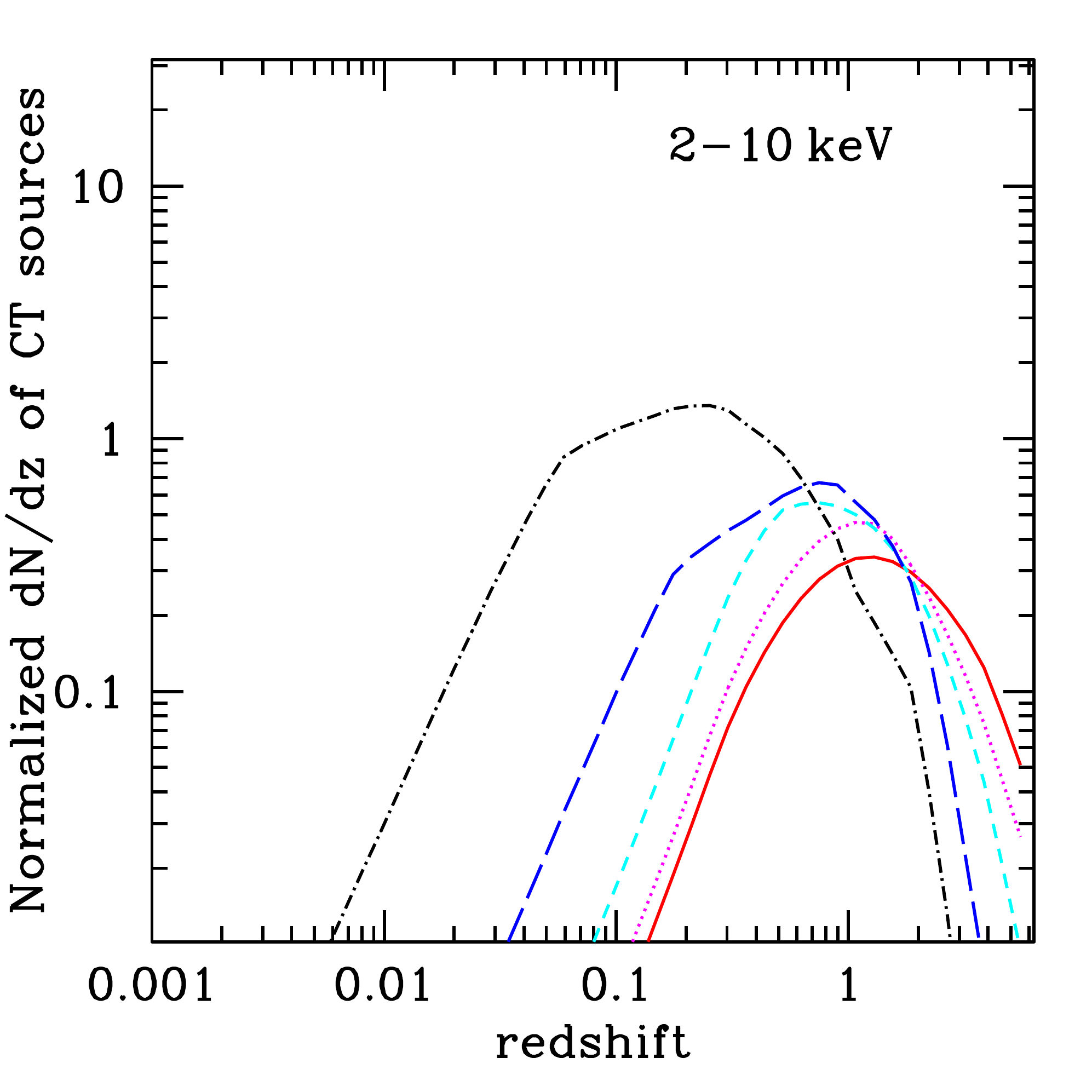}
\includegraphics[height=8cm, width=8cm]{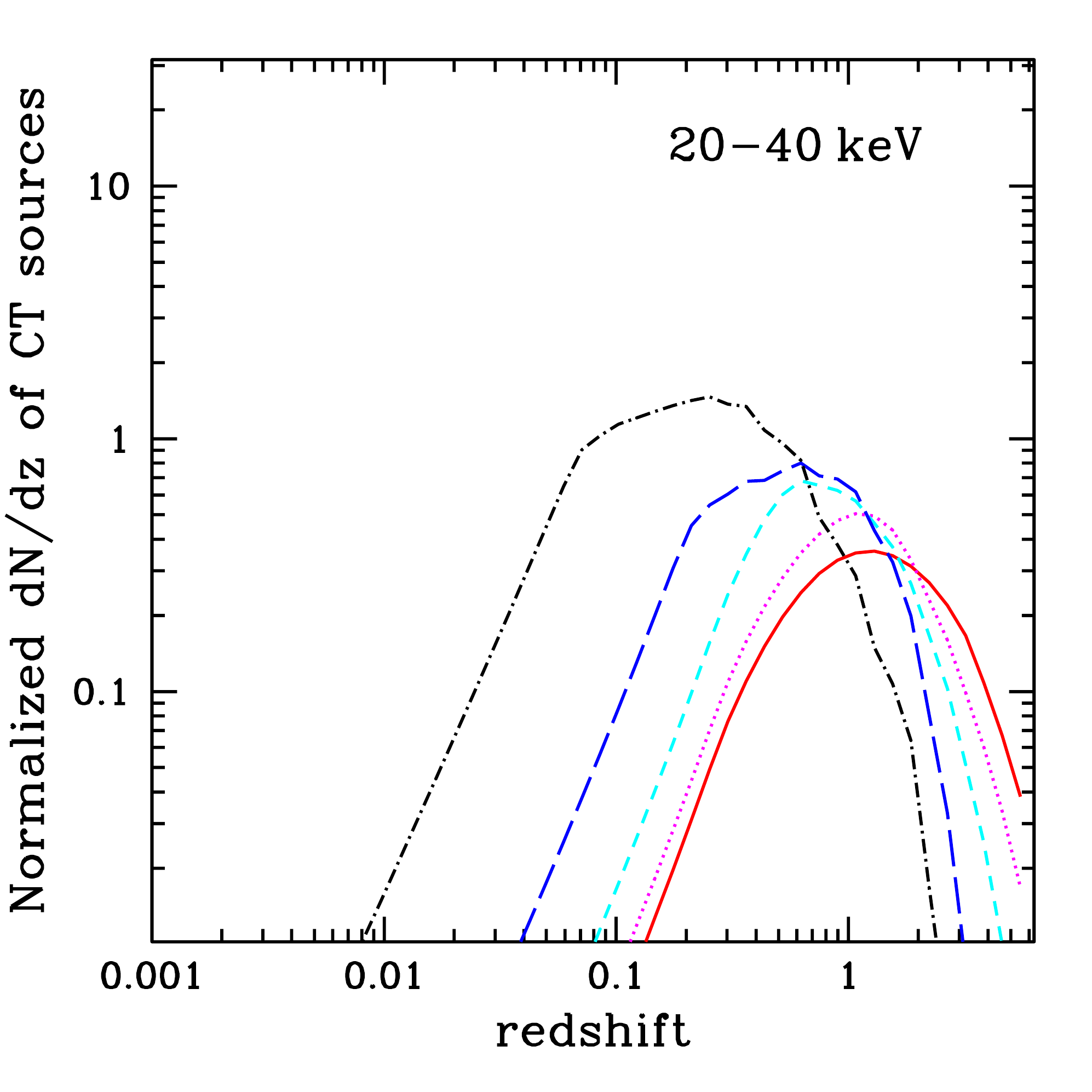}

\caption{The differential  redshift distribution of  Compton-thick AGN
(dN/dz) predicted  by our  model in the  spectral bands  0.5-2, 0.5-8,
2-10 and  20-40\,keV.  The different curves correspond  to flux limits
of  $10^{-13}$ (black  dash-dotted),  $10^{-14}$ (blue  large-dashed),
$10^{-15}$ (cyan small-dashed),  $10^{-16}$ (red dotted) and $10^{-17}
\rm \, erg \, s^{-1} \, cm^{-2}$ (red continuous).  }\label{nz}
\end{center}
\end{figure*}

\begin{table} \centering

\caption{Model parameters ($\Gamma$,  $E_C$, $f_{REF}$, $f_{CT}$) used
to  predict the  fraction and  number counts  of Compton-thick  AGN in
Figure  \ref{ncounts}.  These  parameters yield  XRB spectra  that are
consistent with  the observations at  the 68\% confidence  level. They
are also consistent with the fraction of Compton-thick AGN in the {\it
SWIFT/BAT} sample of  \citet{Ajello2008_survey} at the 95\% confidence
level. }
\label{tab_count}

\begin{tabular}{cccc}

\hline

$\Gamma$ & $E_C$ & $f_{REF}$ & $f_{CT}$ \\ 

\hline \hline 

1.9      & 245&   0.0650 &  0.05    \\

1.9      & 245&   0.0600 &  0.10    \\

1.9      & 240&   0.0575 &  0.15    \\

1.9      & 235&   0.0550 &  0.20    \\

1.9      & 230&   0.0525 &  0.25    \\

\hline

\end{tabular}





\end{table}

\section{Predictions on the properties of Compton-thick AGN}

Current   and  future  \citep[e.g.    NUSTAR,  \citealt{Harrison2010};
  eROSITA,][]{Predehl2011} high-energy missions have among their prime
scientific  goals the  direct identification  of Compton-thick  AGN to
constrain  their cosmological  evolution.  This  section  presents the
predictions  of   our  model  on  the  number   density  and  redshift
distribution  of  Compton-thick  AGN  at different  energy  bands  and
depths.  

Figure \ref{ncounts} plots as a function of flux the observed fraction
and cumulative  number counts  of Compton-thick sources  determined by
our XRB synthesis code for different values of the parameter $f_{CT}$.
The curves in that figure  were constructed by fixing $\Gamma=1.9$ and
then setting $E_{cut}$, $f_{REF}$ to the values that yield the minimum
$\chi^2$   for   the  XRB   spectrum   (see  Table   \ref{tab_count}).
Predictions  are presented in  the 0.5-2,  0.5-8, 2-10  and 20-40\,keV
bands.   The  redshift  distributions  of  the  Compton-thick  AGN  at
different flux  limits in  those bands are  shown in  Figure \ref{nz}.
The  curves  in   that  figure  correspond  to  the   model  of  Table
\ref{tab_count} with $f_{CT}=15$\%.

The 0.5-2\,keV  band is not an  obvious choice of  energy interval for
Compton-thick   AGN  searches.   However,   many  current   (e.g  {\it
  XMM-Newton}, {\it Chandra}) and future (e.g.  eROSITA) missions have
their highest throughput in  that energy range. Therefore, the deepest
current images  of the  X-ray sky, in  terms of resolving  the highest
fraction of the XRB, are in the 0.5-2\,keV band.  The detectability of
Compton-thick AGN in  the 0.5-2\,keV band depends on  both the assumed
$f_{CT}$ and the adopted level  of the soft component (e.g.  scattered
radiation)  of the  AGN spectra.   The latter  parameter has  a strong
impact on  the number  density of Compton-thick  AGN predicted  by our
model.

The choice  of the 0.5-8\,keV band  is motivated by  recent results on
the  number density  of Compton-thick  AGN  in that  spectral band  by
\cite{BrightmanUeda2012}. They used the 4\,Ms Chandra Deep Field South
data    in     combination    with    the     spectral    models    of
\cite{BrightmanNandra2011_model}, which account for Compton-scattering
and  the   geometry  of   the  circumnuclear  material,   to  identify
Compton-thick AGN to the flux limit  $f_X(\rm 0.5 - 8 \, keV ) \approx
10^{-16} \, erg \, s^{-1} \, cm^{-2}$. Figure \ref{ncounts} shows that
the cumulative number counts of  those sources are consistent with the
$f_{CT}=25$\% model  curve. We chose to compare  our model predictions
with   the   observed   cumulative   number  count   distribution   of
Compton-thick  sources  rather than  their  fraction  relative to  the
overall AGN  population.  This  is to avoid  any uncertainties  in the
determination  of the  number density  as a  function of  flux  of the
overall AGN population with $L_X(\rm 2 -  10 \, keV ) > 10^{42} \, erg
\,  s^{-1}$  adopted  by  \cite{BrightmanUeda2012}  to  normalise  the
Compton-thick number counts.  In the context of our model the fraction
$f_{CT}\approx25$\% implied by the \cite{BrightmanUeda2012} results is
consistent at the 95\% confidence  level with the observed fraction of
Compton-thick sources in  the {\it SWIFT/BAT} AGN sample  in the local
Universe   (see  Figure  \ref{fig_contours_bat}).    This  tentatively
suggests  a  redshift  dependence  of  the $f_{CT}$,  given  the  mean
expected   redshift,  $z\approx1$,  of   the  \cite{BrightmanUeda2012}
Compton-thick AGN sample (see Figure \ref{nz}).
      
The 2-10\,keV  band is routinely  used to search for  heavily obscured
AGN.   This is  because  photons  at this  energy  interval are  least
affected by moderate obscuring column densities, while the sensitivity
of  imaging telescopes,  such as  \xmm\,  and {\it  Chandra}, at  this
energy   range  remains   relatively  high.    The   detectability  of
Compton-thick AGN  in the 2-10\,keV band  depends on the  level of the
reflection component of the  AGN spectra.  

The 20-40\,keV band encompasses the  peak of the XRB spectrum and will
be explored to unprecedented depths ($\sim 10^{-14}\rm \, erg\, s^{-1}
\,  cm^{-2}$) with  the imaging  optics  of the  {\it NUSTAR}  mission
\citep[Nuclear  Spectroscopic Telescope  Array,][]{Harrison2010}.  The
Compton-thick AGN fraction predictions  in that band are more directly
related to $f_{CT}$ and are  least dependent on other input parameters
to  the  XRB synthesis  code,  such as  the  level  of the  reflection
component.   At the  confusion flux  limit of  {\it  NUSTAR}, $f_X(\rm
20-40\,keV) \approx 10^{-14} \, erg \, s^{-1} \,cm^{-2}$, we predict a
fraction  of  Compton-thick  AGN  between  2  to  10\%,  depending  on
$f_{CT}$.   Figure \ref{nz}  shows that  the redshift  distribution of
those sources is expected to peak at $z\approx 0.6$.

Although the efficiency of finding Compton-thick AGN is highest in the
20-40\,keV band, the largest numbers  of such sources will be produced
by wide-area  surveys at lower  energies.  At the flux  limit $f_X(\rm
0.5-2)\approx10^{-14} \rm erg \, s^{-1} \, cm^{-2}$ of the eROSITA All
Sky Survey for example,  the predicted number density of Compton-thick
AGN is 0.2-0.9$\rm  deg^{-2}$.  This translates into a  total of 900
to  4000   Compton-thick  AGN  in  the   0.5-2\,keV  eROSITA  source
catalogue.  Figure  \ref{nz} shows that the mean  expected redshift of
these sources is $z\approx 0.3$.   In the 2-10\,keV band, eROSITA will
achieve a sensitivity of about  $3\times 10^{-13} \rm erg \, s^{-1} \,
cm^{-2}$.   A total  of about  70 to  280 Compton-thick  AGN are
expected at that  flux limit over the whole  sky. The relatively small
field of view of {\it NUSTAR}  (6\,arcmin radius) and the fact that it
is  confusion limited  at  $f_X(\rm 20-40)\approx10^{-14}  \rm erg  \,
s^{-1} \,  cm^{-2}$ (about 100\,ks exposure), means  that this mission
will detect  up to a few  hundred Compton-thick AGN,  depending on the
adopted survey strategy. A limitation of the eROSITA data, however, is
that  the  detected sources  will  typically  have  limited number  of
counts.  The identification of the Compton-thick AGN among the eROSITA
X-ray source population will therefore be challenging.

\section{Discussion}
We presented  a synthesis model for  the XRB that uses  the results of
state-of-the-art  Monte   Carlo  simulations  for   the  key  spectral
components of AGN.  This was  used to reconstruct the XRB spectrum for
a wide range of values of the parameters (i) $\Gamma$, the exponent of
the intrinsic AGN power-law, (ii) $E_C$, the high-energy cutoff of the
AGN spectra, (iii)  $f_{REF}$, the level of reflection  in the spectra
of AGN, and (iv) $f_{CT}$,  the fraction of Compton thick sources.  We
then  define the volume  of the  four-dimensional parameter  space for
which the model XRB spectra are consistent with the observed one.

A result from Figures \ref{fig_contours} and \ref{fig_contours_bat} is
that there are no solutions  consistent with the observed XRB spectrum
for  $\Gamma\ga1.98$ (better  than  $95$ per  cent confidence  level),
under  the assumptions of  the model  and for  the range  of parameter
values explored. Therefore,  within the context of the  model, this is
an upper limit to the intrinsic power-law index of AGN.

It is also interesting in Figure \ref{fig_contours} that the intensity
of  the  integrated  XRB  spectrum  does not  provide  any  meaningful
constraints on the space density of Compton-thick AGN in the Universe.
Our analysis  shows that a wide  range of values  of the Compton-thick
fraction, $f_{CT}$, yields acceptable  fits to the XRB spectrum.  This
is because of degeneracies among input parameters to the XRB synthesis
code.  Also the parameters that  describe the shape of the AGN spectra
(e.g.   reflection  fraction,  high   energy  cutoff)  are  still  not
adequately  constrained by  observations  and therefore  the range  of
plausible  values remains  wide.  Figure  \ref{fig_contours_bat} shows
that the most solid limits  on the fraction of Compton-thick AGN comes
from the direct detection of such sources.

This  is also demonstrated  in Figure  \ref{ncounts}, which  plots the
model  expectations for  the  observed fraction  and the  number-count
distribution of Compton  thick AGN as a function  of flux at different
energy bands.  The  number counts of Compton thick  AGN in the Chandra
Deep Field  South (Brightman  \& Ueda 2012)  suggest $f_{CT}\approx25$
per cent  at $z\approx1$, in agreement  at the 95  per cent confidence
level with the Swift/BAT results in the local Universe.

Figure  \ref{ncounts}  also compares  our  predictions  on the  number
density  of  Compton-thick AGN  with  the  X-ray background  synthesis
models                                                               of
\cite{Gilli2007}\footnote{http://www.bo.astro.it/$\sim$gilli/counts.html}
and \cite{Draper_Ballantyne2010}.   The former  model is built  on the
same basic principles (e.g. AGN  unification paradigm) as the XRB code
presented in this paper.   There are important differences between the
two codes however, in e.g.  the adopted X-ray luminosity function, the
overall fraction of Compton-thick AGN, and the level of the reflection
component.   Only mildly  Compton-thick AGN,  $\rm \log  N_H= 24  - 25
\,cm^{-2}$, are  included in the  \cite{Gilli2007} predictions plotted
in Figure \ref{ncounts} to facilitate the comparison with our results.
Contrary  to other  XRB synthesis  codes, \cite{Draper_Ballantyne2010}
postulated  that  the physical  parameter  that  is  more relevant  to
obscuration and Compton-thickness is the accretion rate onto the SMBH.
This is motivated by numerical simulations, which suggest that the AGN
obscuration is related to the  evolutionary status of the SMBH, and by
observational  constraints  on  the  accretion properties  of  nearby
Compton-thick  QSOs.    This  model  assumes  that   the  fraction  of
Compton-thick  AGN ($\rm \log  N_H= 24  - 25  \,cm^{-2}$) is  a strong
function   of   accretion   rate.    They  dominate   at   both   high
($\lambda>0.9$;   $f_{CT}\approx0.86$)    and   low   ($\lambda<0.01$;
$f_{CT}\approx0.60$)  Eddington  ratios,  while  they  are  absent  at
intermediate Eddington ratios.

In  all  panels  of  Figure  \ref{ncounts}  the  \cite{Gilli2007}  and
\cite{Draper_Ballantyne2010}  results predict  a  very different  flux
dependence  of the  observed Compton-thick  fraction and  number count
distribution.   In  the 0.5-8  or  2-10\,keV  bands  for example,  the
differences are  caused by the  adopted $f_{CT}$ and the  treatment of
the reflection spectral component in the different codes.  We assigned
the same $f_{REF}$ to all  AGN, whereas in \citet{Gilli2007} the level
of reflection depends on the obscuration and accretion luminosity.  As
a result, the \citet{Gilli2007}  predictions cut through our curves in
Figure \ref{ncounts}.   At faint fluxes, below about  $10^{-16} \rm \,
erg \, s^{-1} \, cm^{-2}$, their curve lies well above our predictions
because of the  higher fraction of Compton-thick AGN  adopted by their
model.    The   \cite{Draper_Ballantyne2010}   model   predicts   more
Compton-thick  AGN  relative to  the  Gilli  et  al.  (2007)  and  our
predictions.   The hump  in their  number-count predictions  marks the
change in  the dominant  Compton-thick population from  high Eddington
ratio sources at bright fluxes to slowly accreting systems.

The detectability of Compton-thick  AGN in the 0.5-2\,keV band depends
quite strongly on the level of the soft spectral component.  It is not
surprising  that  in  Figure  \ref{ncounts} there  is  fair  agreement
between our $f_{CT}=25$\% model  and that of \citet{Gilli2007} down to
$f_X(\rm  0.5-2\,keV)\approx10^{-16}\,  erg  \,  s^{-1}  \,  cm^{-2}$,
although the  latter assumes a  much higher fraction  of Compton-thick
AGN.   This is because  both models  use the  same level  of scattered
radiation (3\%) to model the soft  component of AGN. The upturn in the
fraction of the Compton-thick AGN at $f_X(\rm 0.5-2\,keV) \la 10^{-16}
\rm \, erg \, s^{-1} \, cm^{-2}$  in the Gilli et al. (2007) models is
caused  by the increasing  contribution of  high redshift  sources for
which the reflection component is redshifted into the 0.5-2\,keV band.

Finally,  the detection rate  of Compton-thick  AGN in  the 20-40\,keV
band is  more directly related to  the total number of  sources in the
parent population  and less sensitive to  the spectral characteristics
of AGN  (e.g.  level of  scattered radiation or  reflection fraction).
Our estimates for  the observed fraction of Compton-thick  AGN in that
band are  therefore consistently below those  of \citet{Gilli2007} and
\cite{Draper_Ballantyne2010}  because these  authors  assume a  higher
intrinsic fraction of Compton-thick AGN.

\section{Conclusions}

A model  for the  synthesis of  the XRB was  presented, which  used as
input  an  AGN  X-ray   spectral  library  generated  by  Monte  Carlo
simulations  to  account  for  the impact  of  Compton-scattering  and
photoelectric   absorption.  We  then   explored  whether   the  input
parameters  to   the  XRB  synthesis   code  can  be   constrained  by
observations  of the  XRB intensity  as  a function  of energy.   Four
parameters  were  used  in  this  exercise, the  power-law  index  and
high-energy cutoff of  the intrinsic X-ray spectrum of  AGN, the level
of the reflection component in AGN spectra, and the intrinsic fraction
of Compton-thick  AGN.  We mapped the volume  of this four-dimensional
space that is consistent  with observational determinations of the XRB
spectrum in the  energy interval 3-100\,keV and found  that one of the
least constrained  model parameters  is the fraction  of Compton-thick
AGN  in  the Universe.  Acceptable  fits  to  the XRB  spectrum  (68\%
confidence level)  can be obtained for Compton-thick  AGN fractions in
the range 5--50\%.  This is  because of aliases among input parameters
to the XRB  synthesis code as well as because  of uncertainties in our
current understanding of AGN spectra (e.g.  level of reflection).  The
most stringent constraints on the fraction of Compton-thick AGN in the
Universe  comes   from  the  direct  detection  of   such  sources  in
high-energy surveys, such as those  carried out by {\it SWIFT/BAT} and
{\it INTEGRAL}.  These  observations suggest a Compton-thick fraction,
at least  in the  local Universe, of  10--20\% at the  68\% confidence
level.  We also predicted the  number counts of Compton-thick AGN as a
function of flux and energy  band.  These predictions can be tested by
current and  upcoming X-ray missions  to set tight constraints  on the
space density of Compton-thick AGN as a function of redshift.

\begin{acknowledgements}

We  thank   the  anonymous  referee  for   constructive  comments  and
suggestions.  AG acknowledges financial  support from  the Marie-Curie
Reintegration Grant PERG03-GA-2008-230644.  IG acknowledges support by
the   European   Community   through   the  Marie   Curie   fellowship
FP7-PEOPLE-IEF-2008 Prop. 235285 under the Seventh Framework Programme
(FP7/2007-2013).

\end{acknowledgements}

\bibliography{xray}{} \bibliographystyle{aa}

\end{document}